\documentclass[aps,preprint,showpacs,superscriptaddress,footinbib,groupedaddress]{revtex4-1}

\usepackage{dsfont}
\usepackage{amsmath}
\usepackage{graphicx}
\usepackage[utf8]{inputenc}
\usepackage{ulem} 
\usepackage{units} 
\usepackage{version} 
\usepackage{color}
\usepackage{colortbl}
\usepackage{xcolor}  
\usepackage{fancyref}
\usepackage[colorlinks=true,citecolor=blue,urlcolor=blue,linkcolor=blue]{hyperref} 

\usepackage{multirow}
\usepackage{array}
\usepackage{booktabs}
\usepackage{longtable}

\newcommand{\ms}{\ \text{ms}}

\begin{document}

\title{Improving the phase response of an atom interferometer by means of temporal pulse shaping}

\author{Bess Fang}
\email{bess.fang@obspm.fr}
\author{Nicolas Mielec}
\author{Denis Savoie}
\author{Matteo Altorio}
\author{Arnaud Landragin}
\author{Remi Geiger}
\email{remi.geiger@obspm.fr}

\affiliation{LNE-SYRTE, Observatoire de Paris, PSL Research
  University, CNRS, Sorbonne Universit\'es, UPMC Univ. Paris 06, 61
  avenue de l'Observatoire, 75014 Paris, France}

\date{\today}

\begin{abstract}
\label{abstract}
We study theoretically and experimentally the influence of temporally shaping the light pulses in an atom interferometer, with a focus on the phase response of the interferometer. We show that smooth light pulse shapes allow rejecting high frequency phase fluctuations (above the Rabi frequency) and thus relax the requirements on the phase noise or frequency noise of the interrogation lasers driving the interferometer.  The light pulse shape is also shown to modify the scale factor of the interferometer, which has to be taken into account
in the evaluation of its accuracy budget. We discuss the trade-offs to operate when choosing a particular pulse shape, by taking into account phase noise rejection, velocity selectivity, and applicability to large momentum transfer atom interferometry.
\end{abstract}



\maketitle

\section{introduction}
Precision measurements rely on a careful analysis of the relevant
noise sources and systematic effects.  In the field of inertial
sensors, instruments based on light-pulse atom interferometry allow
measurements of gravito-inertial effects such as linear accelerations
\cite{Canuel2006,Geiger2011,Rakholia2014}, rotations
\cite{Gustavson2000,Tackmann2012,Dutta_2016}, Earth gravity field \cite{LouchetChauvet2011,Freier2016} and
of its gradient \cite{McGuirk2002} or curvature \cite{Rosi2015}. They
have also been used for precise determinations of fundamental
constants \cite{Bouchendira2011,Rosi2014} and tests of the weak
equivalence principle (see,
e.g. \cite{Varoquaux2009,Schlippert2014,Zhou2015,Bonnin2013,Duan2016,Aguilera2014,Barrett2016,Rosi2017,Overstreet2017}),
and have been proposed for gravitational wave detection in the sub-10
Hz frequency band \cite{Dimopoulos2009,Geiger2016}.  These sensors most often use two counter-propagating laser beams to realize the beam
splitters and mirrors for the atomic waves associated to two different momentum states.  The  stability and accuracy
of the sensors critically depends on the control of the intensity and of the relative   phase of these two lasers, both spatially and temporally.  
For example, the spatial profile of the relative laser  phase is the main source of systematic
effects in most accurate atomic gravimeters
\cite{LouchetChauvet2011,Freier2016}, and is an important concern in
the design of future gravitational wave detectors based on atom
interferometers (AIs) \cite{Hogan2011}.

The temporal shape of the light-pulses (i.e. of the laser intensity) driving an AI
determines the  efficiency of the beam splitters and mirrors acting on the two momentum states of the AI.  More
precisely, for velocity selective transitions, the transfer efficiency
of the pulse is given by the convolution between the velocity
distribution of the atoms and the Fourier transform of the pulse
shape \cite{Kasevich1991vel}.  
Efficient transitions (i.e. high contrasts) can thus be achieved by temporally shaping the pulse intensity and
phase, as shown in~\cite{LuoYukun2016,Dunning2014}.
  Moreover, when
driving an interferometer with large momentum transfer (LMT) atom
optics, it has been shown that  pulses of Gaussian temporal shape
significantly improve the transfer efficiency with respect to
rectangular pulse shapes \cite{Muller2008a,Szigeti2012}.  
Adiabatic rapid passage (see, e.g. \cite{Pereira_Dos_Santos_2002}) was also considered
in LMT inteferometry, but was shown to require stringent control of the laser phase noise compared to pulse shaping \cite{Kovachy_2012}.

  In addition to the influence on the contrast of the interferometer, the temporal shape of the pulse is expected
to affect the (frequency-dependent) response of the interferometer to
phase fluctuations, which is an important source of instability in AIs.  Furthermore, as the phase
response of the AI is modified, pulse shaping should introduce a
correction to the scale factor of the interferometer, which has to be
accounted for in the accuracy budget of atomic sensors.

In this article, we  study   the phase response of an AI driven by arbitrary temporal  light pulse shapes.
Our main interest is to highlight the strong difference in the phase response of an AI driven by rectangular and smooth pulse shapes.
We concentrate on a few pulse shapes that are representative for the optimization of the following criteria: rejection of high-frequency  laser phase (or frequency) noise, velocity selectivity of the pulse, and applicability to LMT interferometry.
Experimentally, we focus on the comparison of the phase sensitivity function (section \ref{sec:goft}) and of the rejection of laser phase noise (section \ref{sec:noise}) between the two mostly employed rectangular and Gaussian pulses, in order to validate our calculations. 
In addition to the rectangular and Gaussian pulses, we   discuss two other representative pulse shapes: \textit{(i)}  the GSinc pulse, which is  the product
of a Gaussian and a cardinal sine, and \textit{(ii)} the Gaussian-Flat pulse (labelled GFlat thereafter) which is a flat pulse with gaussian edges.
For completeness of the presentation, we study in section \ref{sec:selectivity} the influence of pulse shaping on the frequency selectivity, in line with previous works \cite{LuoYukun2016,Dunning2014}.
Finally, we present in section \ref{sec:scale_fact} a correction to the interferometer scale factor associated with pulse shaping, and discuss its relevance for different precision measurements involving AI based  sensors. 
We conclude our paper with a discussion  of the trade-offs to operate when  selecting a given pulse shape for a particular application  (section \ref{sec:discussion}).


\section{sensitivity function with arbitrary pulse shape}\label{sec:goft}

\subsection{Theory}

The sensitivity function was first introduced to study the degradation
of an atomic clock due to the phase noise of the local
oscillator~\cite{Dick_1987}, but the idea is more general.  It
describes the response of an atom interferometer phase to
infinitesimal changes of external parameters.  We investigate here the
response of the AI phase $\delta \Phi$ to an instantaneous variation
$\delta \phi(t)$ of the relative phase between the two lasers driving
the AI, occurring at a given time $t$. As in previous works
\cite{Cheinet_2008}, we define the sensitivity function as
\begin{equation}
  g(t) = \lim_{\delta \phi \rightarrow 0} \frac{\delta \Phi (\delta
    \phi, t)}{\delta \phi(t)}.
\end{equation}
It can be calculated for an interferometer composed of perfect
beam-splitters and mirrors using
\begin{equation}
g(t) = \sin\left(\int_{t_0}^t \Omega(t') dt'\right),
\label{eqn:integral_g}
\end{equation}
where $\Omega(t)$ is the Rabi frequency seen by the atoms during the
interferometric sequence~\cite{BizePhD}, with $\Omega(t)=0$  for  $t<t_0$.  The overall shape of $g(t)$
depends on the AI configuration, i.e. on the number of light-pulses.
In this work, our main interest lies in the effect of temporal pulse
shape.  Therefore we consider without loss of generality, a two-light
pulse interferometer, i.e. a Ramsey configuration.  For a Ramsey
sequence with two rectangular $\pi/2$ pulses characterized by a rabi
frequency $\Omega_R$ and duration $\tau$ separated by Ramsey time $T$,
the sensitivity function reads
\begin{equation} \label{eqn:goft}
g(t) = \left\{ \begin{array}{lll}
    \sin\big(\Omega_R \times (t + \mathrm{\frac{\tau}{2}}) \big) & {\rm for} & -\mathrm{\frac{\tau}{2}}<t<\mathrm{\frac{\tau}{2}}\\
    1 & & \mathrm{\frac{\tau}{2}}<t<T-\mathrm{\frac{\tau}{2}}\\
    \sin\big(\Omega_R \times (t-T+\mathrm{\frac{3\tau}{2}})\big) & & T-\mathrm{\frac{\tau}{2}}<t<T+\mathrm{\frac{\tau}{2}}
    \end{array} \right. \!\!,
\end{equation}
where the origin of the time axis is (arbitrarily) aligned with the
center of the first light pulse.

\begin{figure}[!bth]
\centering
\includegraphics[width=0.7\linewidth]{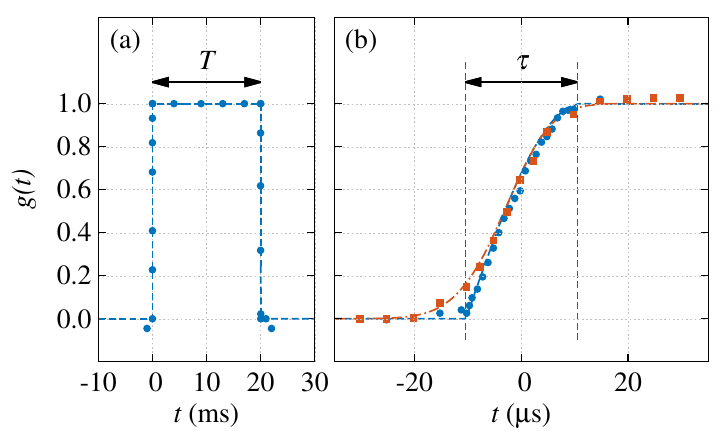}
\caption{\label{fig:sensitivity} Sensitivity function $g(t)$ of a
  Ramsey sequence.  (a) Complete $g(t)$ for two rectangular $\pi/2$
  pulses separated by Ramsey time $T$. (b) Zoom on the rising slope for
  rectangular (blue) and Gaussian (red) pulses.  We compare
  calculations (lines) according to Eq.~(\ref{eqn:goft}) and our
  measurements (points). Error bars on the measurements are smaller
  than the plot symbol.}
\end{figure}

We show $g(t)$ as a dashed line in Fig.~\ref{fig:sensitivity}~(a).  In
the limit of infinitely short laser pulses, $g(t)$ is box-like, as the
interferometer copies the phase jitter of the interrogation laser
($g(t)=1$) between the two laser pulses.

We show in Fig.~\ref{fig:sensitivity}~(b) a zoom of of the rising
slope (i.e. during the first $\pi/2$ pulse) of $g(t)$ for a sequence
based on rectangular pulses (blue dashed line) and Gaussian pulses
(red dash-dotted line).  We have chosen to use the same peak intensity
in our calculation (and our experiments later), and adjust the pulse
duration to obtain the desired Rabi angle.  This is motivated by the
fact that the peak laser intensity depends on the total power
available, which is often the limiting experimental factor.  The main
difference in the sensitivity function takes place around $t=-\tau/2$,
where $\tau$ denotes the duration of the rectangular $\pi/2$ pulse.
The sudden intensity variation of a rectangular pulse gives rise to a
fast rise in the sensitivity function, and a discontinuity in its
derivative.  This fast rise is in contrast with the gradual change
induced by a smooth intensity variation of a Gaussian pulse.  Such a
difference results in different spectral behaviors of $g(t)$ for the
two pulse shapes, as we will discuss in Sec.~\ref{sec:noise}.

\subsection{Measurement of the sensitivity function for rectangular and Gaussian pulses}
\label{subsec:measurement_g}

We measure $g(t)$ using the experimental setup described
in~\cite{Meunier_2014,Dutta_2016}.  Briefly speaking, we use an atomic
fountain to prepare cold cesium-$133$ atoms.  At each experimental
cycle, about $10^6$ atoms are prepared into the magnetically
insensitive $|F=4,~m_F=0\rangle$ ground state and launched into the
interferometric zone.  The Ramsey pulses are realized via stimulated
Raman transitions, using a doubly seeded tapered
amplifier~\cite{L_v_que_2010}.  The seeding external cavity diode
lasers have a fixed phase relation by means of an optical phase locked
loop (PLL) close to the Cs clock transition frequency.  Both lasers
are about $500$~MHz red-detuned from the excited state of the ${\rm
  D}_2$ line to reduce spontaneous emissions during the Raman
transition.  At the end of the interferometer sequence, the population
in each of the hyperfine ground states $N_3$ and $N_4$ is detected by
fluorescence, and the transition probability is obtained by $P_4 =
N_4/(N_3+N_4)$.

\begin{figure}[!bth]
\centering
\includegraphics[width=0.7\linewidth, trim = 4cm 6.5cm 2.5cm 3.5cm,
    clip]{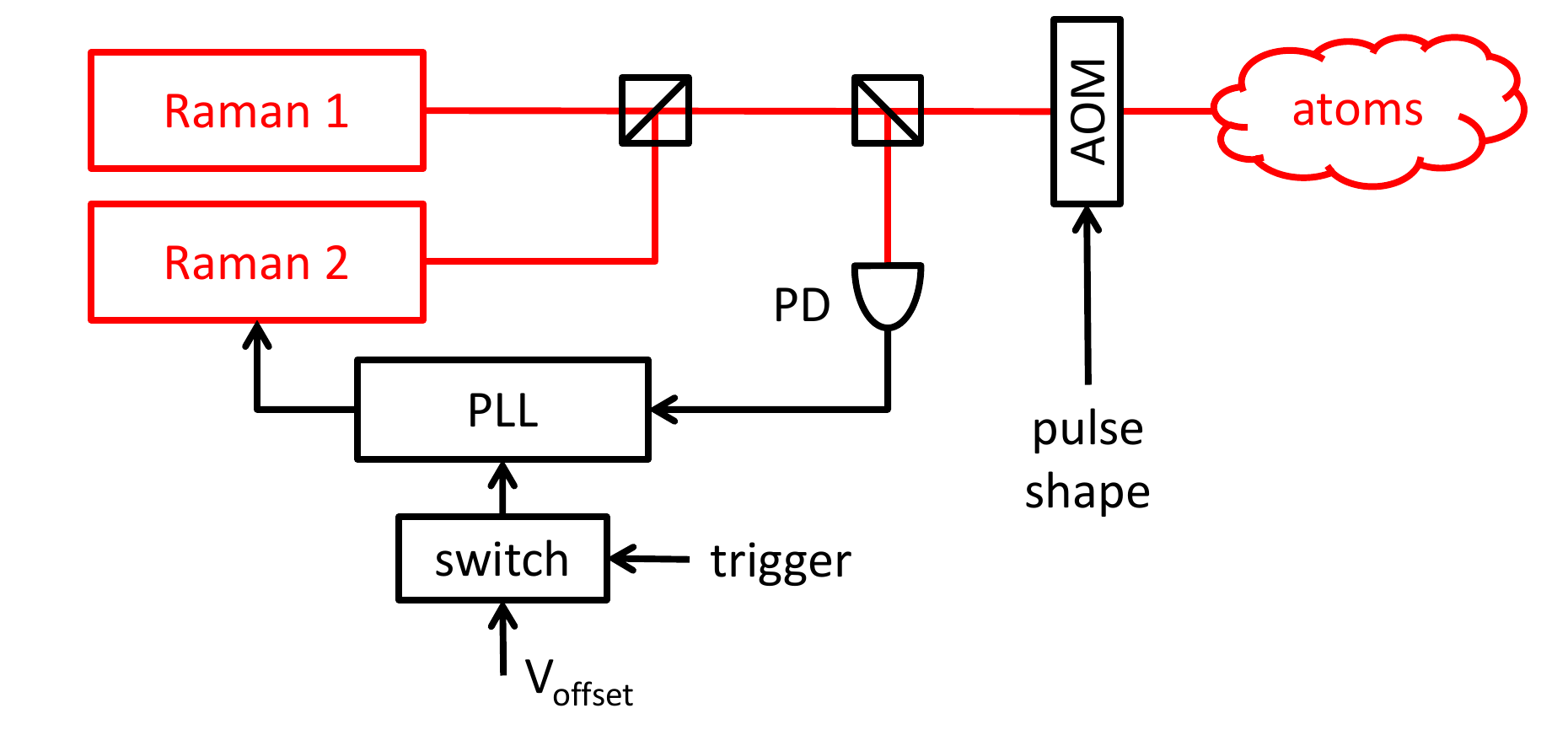}
\caption{\label{fig:phasejump} Schematic of the phase jump control.
  The beat note of two lasers (Raman 1 and 2) is detected on a fast
  photodiode (PD) and phase locked onto a reference signal at the Cs
  ground-state hyperfine splitting frequency of about $9.192$~GHz.
  Phase jumps are implemented by sending a DC voltage $V_{\rm offset}$
  to the feedback port of the PLL through a voltage controlled switch.
  By appropriately attenuating the radio-frequency signal driving the
  AOM, arbitrary temporal profiles of laser pulses can be sent onto
  the atoms.  }
\end{figure}

The laser phase jump is implemented by applying a DC voltage $V_{\rm
  offset}$ to the feedback port in the PLL through a voltage
controlled switch, which is triggered at different times.  See
Fig.~\ref{fig:phasejump} for the control schematics.  The voltage
offset corresponds to a phase jump of about $340$~mrad.  The switch
has a delay of $0.3~\mu$s, where as the PLL has a locking bandwith of
$1.6$~MHz. Thus, the total delay in the phase jump implementation is
under $1~\mu$s, much shorter than the duration of the rectangular
$\pi/2$ pulse $\tau=21~\mu$s (peak Rabi frequency $\Omega_R/2\pi =
12$~kHz).

We shape the Raman light pulses by attenuating the radio-frequency
signal driving the acousto-optic modulator (AOM), which controls the
intensity of the Raman pulses shone on the atoms.  A commercial direct
digital synthesizer (Rigol 4620) is used to generate a wave form that
takes into account the desired wave form (e.g. a Gaussian pulse) as
well as the response of the chain of a voltage-controlled attenuator
followed by an RF amplifier.  This response is calibrated against a
monitor photodiode in order to ensure that the intensity of the Raman
pulses follows the desired wave form.

With a Ramsey time of $T=20$~ms, the phase noise of the clock sequence
is about $30$~mrad.Hz$^{-1/2}$, which enables a mid-fringe operation of the
interferometer.  We further stabilize the phase offset of the
interferometer by applying a mid-fringe lock~\cite{Merlet_2009}, which
converts the measurement of the atomic transition probability directly
to the interferometric phase.  This technique is immune to variations
in the probability offset and reduces the sensitivity to the noise in
the fringe amplitude, thereby allowing a robust measurement of the
interferometric phase.

To compare the experimental data with the calculations, we offset the
measured phase shift to $0$ and normalize by $340$~mrad to obtain the
experimental $g(t)$.  We display our measurements in
Fig.~\ref{fig:sensitivity} (a) for the complete $g(t)$ with
rectangular pulses \cite{Cheinet_2008}.  Fig.~\ref{fig:sensitivity}
(b) shows the rising slope for rectangular (circles) and Gaussian
(rectangulars) pulses.  The relative phase uncertainty of each measurement
is below 4~mrad, i.e. smaller than the plot symbol.  The time axis for
the experimental data is shifted by $0.22~\mu$s to account for the
delay through the switch and the PLL.  Our measurements confirm the
temporal form of $g(t)$ given by Eq.~(\ref{eqn:goft}), and well
resolve the differences between the two pulse shapes implemented.


\section{Frequency response of the AI to pulse shaping}\label{sec:noise}
\subsection{Calculations}

The impact of the sensitivity function on the interferometer phase
noise can be more easily understood in Fourier space.  According
to~\cite{Cheinet_2008,Meunier_2014}, the variance of the interferometric phase
noise can be expressed as
\begin{equation}
  \sigma_\Phi^2 = \int_0^\infty \frac{d\omega}{2\pi}~|H(\omega)|^2
  S_\phi(\omega),
  \label{eq:sigma_phi}
\end{equation}
where the transfer function $H(\omega) = \omega \big| G(\omega)\big|$,
$G(\omega)$ is the Fourier transform of the sensitivity function
$g(t)$, and $S_\phi(\omega)$ is the power spectral density of the
Raman laser phase noise.

\begin{figure}[!bth]
\centering
\includegraphics[width = 0.9\linewidth]{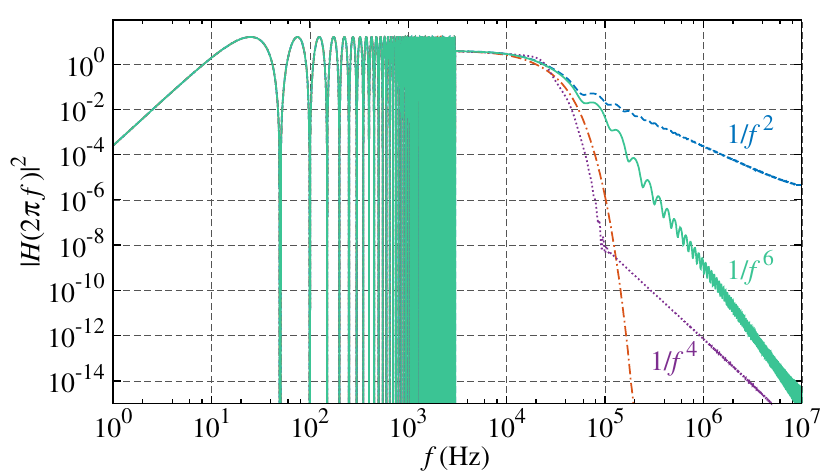}
\caption{\label{fig:Homega} Transfer function $|H(2\pi f)|^2$ for a 3 pulse AI with $T=20$~ms driven by different pulse shapes:  rectangular
  (blue dashed line), Gaussian (red dash-dotted line), GSinc (purple
  dotted line), and GFlat (green solid line). The peak Rabi frequency is the same for all pulse shapes.}
\end{figure}

We plot in Fig.~\ref{fig:Homega} the transfer function $|H(2\pi f)|^2$
as a function of frequency $f$ for a 3 light pulse sequence ($\pi/2-\pi-\pi/2$) for various pulse shapes:  rectangular
(blue dashed line), Gaussian (red dash-dotted line), GSinc (purple
dotted line), and GFlat (green).  
The peak Rabi frequency is the same for all pulse shapes.
The calculation is analytic for rectangular
pulse and numerical for the other pulse shapes. 
The Gaussian pulse is truncated at 6 standard deviations on both sides. 
The definition of the GSinc and GFlat pulse shapes is given in appendix \ref{sec:pulsedef}. 

Independent of the pulse shapes used, the transfer function $|H(2\pi
f)|^2$ is oscillatory with arches spanning $1/T$, i.e.\ $50$~Hz for
our choice of $T = 20$~ms.  This is illustrated at low frequency up
to $3$~kHz, beyond which we plot the mean value over $3$~kHz in
order to illustrate the general frequency dependence of the envelope.  

The difference between the four pulse shapes lies mainly in the
low-pass cut-off occurring near the peak Rabi frequency (here
$12$~kHz).  For a rectangular pulse, the high-frequency noise is filtered out with a $1/f^2$ scaling of $H^2$, whereas the use of smoother pulses warrants a significantly faster decay, and therefore a better suppression of high-frequency noise.  In particular, Gaussian pulses give rise to the strongest high-frequency cut-off in $H^2$.  
The GSinc pulse gives a similar behavior as the Gaussian pulse around the peak Rabi frequency, before following a $1/f^4$ scaling at high frequency. The frequency at which the slope changes  is determined by   the width of the Gaussian relative to the length of the sine cardinal (the smaller the width of the Gaussian, the further the change of slope).
The GFlat pulse gives rise  to $1/f^6$ scaling in $H^2$ beyond the peak Rabi frequency.  

To understand the asymptotic behavior of the transfer function qualitatively, we performed  calculations with  various pulse shapes, including temporally asymmetric pulses, and using different  shapes for the $\pi/2$ and $\pi$ pulses. 
We found that the high frequency behavior is first determined by the steepness of $g(t)$ at the beginning of the first $\pi/2$  and the end of the last $\pi/2$ pulses. Even faster decay of the transfer function is then related to the steepness of $g(t)$ at the end of the first $\pi/2$ pulse, the beginning of the last $\pi/2$ pulse, and the $\pi$ pulse. Further details on this qualitative interpretation in line with Eq.\eqref{eqn:integral_g} can be found in appendix \ref{sec:qualitative_interpretation}.

\subsection{Measurements of the transfer function}
\begin{figure}[!bth]
\centering
\includegraphics[width = \linewidth]{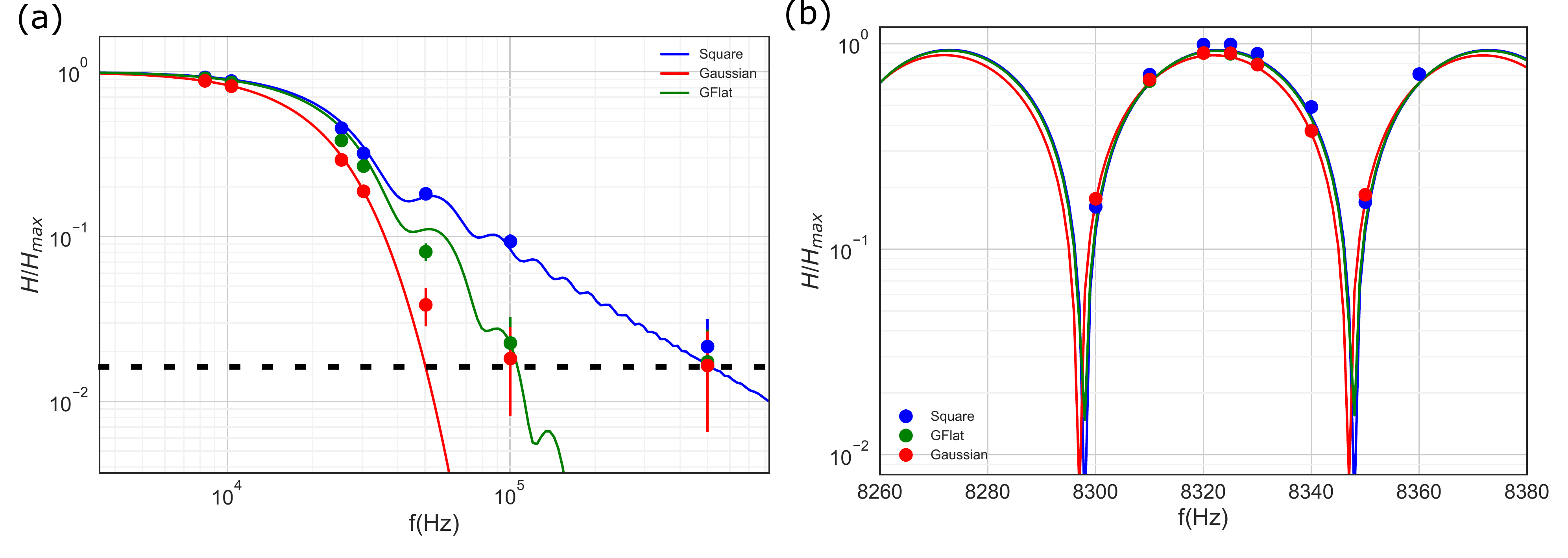}
\caption{\label{fig:Homega_exp} Transfer functions for a Ramsey sequence $\pi/2-\pi/2$ with a Rabi frequency of 8.3~kHz, and a Ramsay time of 20~ms. Three pulse shapes are considered: rectangular (total duration of $30 \ \mu$s), Gaussian, and GFlat. (a) Asymptotic behavior of $H(\omega)$, where the experimental and theoretic data are the maxima of the arches.(b) A zoom around the Rabi frequency.  The errorbars correspond to   statistical errors at the $68\%$ confidence interval. 
The dashed horizontal line in (a) corresponds to the noise floor of our measurements.}
\end{figure}

We measure the transfer function $H(\omega)$ for different pulse shapes by realizing a Ramsey sequence ($\pi/2-\pi/2$) using co-propagating Raman transitions, with a Ramsey time of $T=20 \ms$, and a Rabi frequency of $8.3$~kHz. 
To measure the transfer function, we follow the approach of \cite{Cheinet_2008}: we apply  a sinusoidal phase modulation of angular frequency $\omega$ starting at the first Raman pulse  and lasting during the whole interferometer, and measure its effect on the phase of the atom interferometer.
We perform two measurements corresponding to two quadratures of the phase modulation, which are added quadratically in order to  extract the value of $H(\omega)$.
The maximum of $H(\omega)$ corresponds to a phase shift of 1.05 rad.
The relative uncertainty of the phase measurements are at the level of  $1\%$. 
To show the asymptotic behavior of $H(\omega)$, we  measure the position of the maxima of the transfer function over several decades. 
 The measurements are shown in Fig.~\ref{fig:Homega_exp}(a) , together with the calculation presented in the previous subsection, without free parameters.
 The experimental data and the calculation agree well within the uncertainties of the experimental parameters ($\sim 10\%$ on the Rabi frequency and $\sim 10 \%$ on the position of the maxima at frequencies above 10 kHz).
 In particular, the measurements resolve the difference in asymptotic behavior of the three pulse shapes.
 We also observe that the positions of the zeros of the transfer function  are indistinguishable for all pulse shapes at frequencies lower than the Rabi frequency, as illustrated around 8.3 kHz in panel (b). 

\subsection{Experimental demonstration of noise rejection}
\begin{figure}[!bth]
\centering
\includegraphics[width=0.7\linewidth]{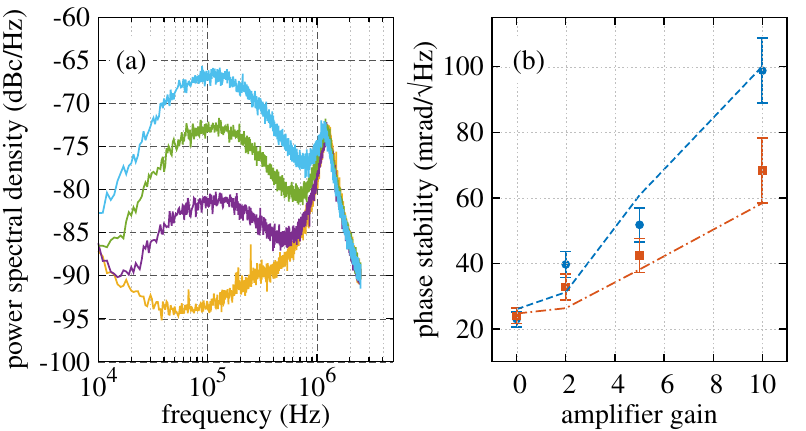}
\caption{\label{fig:filterednoise} (a) Power spectral density of the
  laser phase noise recorded with a spectral analyser with a
  resolution bandwith of $1$~kHz.  The yellow line shows the spectrum
  without additional noise (gain $=0$ in the low-noise amplifier),
  whereas the purple, green and cyan lines correspond to increasing
  noise levels (amplifier gain $=2$, $5$, and $10$).  (b) Short-term
  phase stability of a Ramsey interferometer.  We overlay our
  measurements (points) with calculations (lines) for rectangular
  (blue) and Gaussian (red) pulses.  The errorbars correspond to
  statistical errors at the $68\%$ confidence interval.  }
\end{figure}

To demonstrate experimentally the robustness of smooth pulses
against high-frequency laser phase noise (compared to rectangular pulses), we realize Ramsey sequences ($\pi/2-\pi/2$)
with additional relative phase noise in the Raman lasers.
The difference between the Ramsey sequence and the 3-pulse sequence ($\pi/2-\pi-\pi/2$) only lies in the low frequency behavior of the transfer function (at $f\sim 1/T$), while the high frequency behavior (for $f$ on the order of and higher than the Rabi frequency) is the same for both sequences.
 We concentrate on the comparison between Gaussian and rectangular pulse shapes.
  Adding phase noise is achieved by
sending a noisy signal (instead of a switchable DC voltage as
illustrated in Fig.~\ref{fig:phasejump}) into the feedback port of the
PLL.  We generate a white noise using a commercial synthesizer, filtered into the $40$~kHz to
$300$~kHz band pass and amplified using a commercial low-noise
amplifier.  By varying the amplifer gain, we control the
additional phase noise of the Raman lasers, giving rise to the power
spectral density shown in Fig.~\ref{fig:filterednoise}~(a).  For each
noise level, we measure the short-term phase stability of a Ramsey
sequence ($T=20$~ms) with rectangular (circles) and Gaussian (rectangulars)
pulses, as shown in Fig.~\ref{fig:filterednoise}~(b).  In comparison,
Gaussian pulses consistently rejects a significant fraction of the
additional noise.

We calculate numerically the induced phase noise according to
Eq.~\eqref{eq:sigma_phi}, by numerically integrating over the $10$~kHz
to $2.5$~MHz range. The contribution of frequencies out of this band
is negligible. The total noise is $\sigma
=\sqrt{\sigma_{det}^2+\sigma_{\Phi}^2}$, where $\sigma_{det}=
22$~mrad.Hz$^{-{1/2}}$ is our measured detection noise.  To account
for the uncertainty in the absolute phase noise level applied to the
interferometer, we multiply the phase noise PSD of
Fig.~\ref{fig:filterednoise} (a) by a global factor. This factor is
obtain by matching the calculation and measurement for upper right
point in Fig.~\ref{fig:filterednoise} (a), which is almost exclusively
influenced by phase noise (and not detection noise). Apart from this
global factor common to both pulse shapes, there are no free
parameters.  The calculation follows well the experimental data, and
shows how the Gaussian pulse rejects the high frequency phase noise,
above the Rabi frequency.

\subsection{Discussion and applications to inertial sensors and optical clocks}

The strong rejection of the relative laser phase noise by a smooth pulse
(Gaussian, GSinc, GFlat) at frequencies higher than the Rabi frequency will
help designing optical PLLs for AI experiments, as it relaxes the
requirements on the PLL bandwith.  
Regarding  the limitation to the sensitivity of  cold atom gravimeters due to Raman laser phase noise, we calculate the noise rejection in state of the art instruments. For the work presented in Ref.~\cite{Hu2013}, we compute a phase noise of 7.5 mrad per shot (assuming a $\pi$ rectangular pulse with a duration of  $15 \ \mu$s), in agreement with the measured  short term stability. Using a GFlat pulse yields a noise of 6.1 mrad, and a Gaussian pulse reduces this contribution to 5.9 mrad per shot. 
For the work presented in Ref.~\cite{Le_Gou_t_2008}, the rectangular pulse corresponds to a phase noise of 1.1 mrad per shot, which will be reduced to 0.5 mrad per shot when using GFlat or Gaussian pulses.

In AIs driven by Bragg diffraction, the relative phase noise between the two Bragg lasers is not a
concern, since the two momentum states used in the two interferometer
arms correspond to the same internal energy state. However, because of
propagation delay from the atoms to the mirror which retro-reflects
the Bragg lasers, the laser frequency noise converts into phase noise
on the AI \cite{Le_Gou_t_2007}. Such noise is a major concern in long
baseline AI gradiometers, e.g. in gravitational wave detectors based
on AIs \cite{Geiger2016,Canuel_2014}.  Smooth pulses can therefore relax the
requirements on the laser frequency noise at high frequencies (above
the Rabi frequency, i.e. above typically 10 to 100 kHz).


We also investigate the potential interest of temporally shaping pulses to improve the stability of optical clocks.
The stability of optical clocks critically depends on the frequency stability of the interrogation laser~\cite{quessada2003}, the design of which requires careful attention~\cite{Nicholson2012}.
In that context, we found that  pulse shaping in clocks is less interesting than in AIs.
The reason is that the relevant transfer function for the measurement of frequency (instead of phase) is $|G(\omega)|^2=|H(\omega)|^2/\omega^2$, which scales as $\omega^{-4}$ (for a rectangular  pulse) after the cutoff given by the pulse Rabi frequency $\Omega_0$.
For white frequency noise, the contribution of  high frequencies  ($\omega>\Omega_0$)  is thus $1/3$ of that of  low frequencies ($\omega<\Omega_0$), in power of the noise.
Therefore, faster decay (than $\omega^{-4}$) of the transfer function does  not  significantly impact the stability.
Pulse shaping can however be used to relax the constraints on potential spurious high frequency noise components in the clock laser, e.g. in field applications or compact clock design~\cite{koller2017}.

\section{Frequency selectivity of the pulse}
\label{sec:selectivity}

\begin{figure}[!h]
 \includegraphics[width= \linewidth]{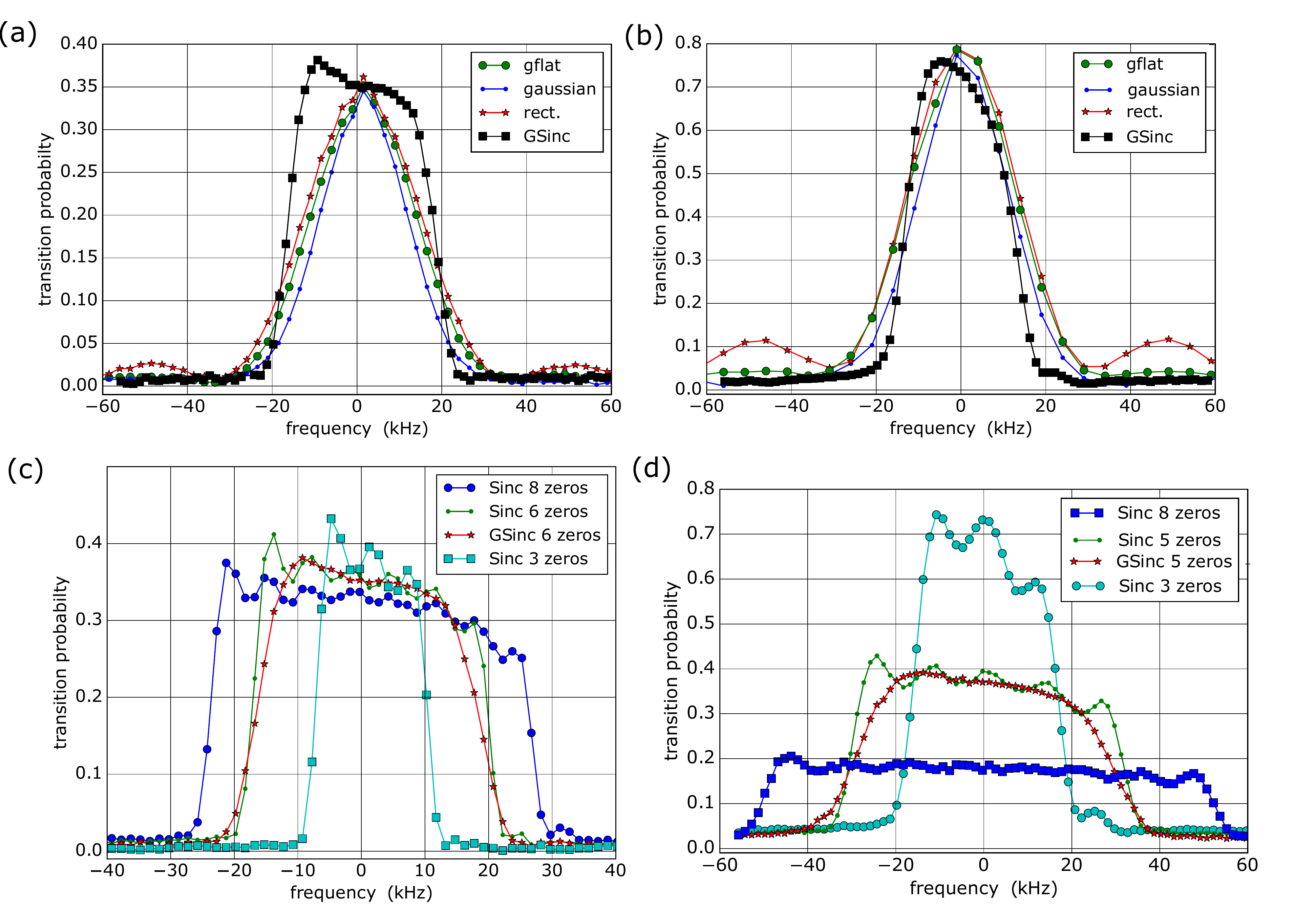}
\caption{\label{fig:spectroscopy}
Spectroscopy of different pulse shapes.
(a): $\pi/2$ pulse.
(b):  $\pi$ pulse.
In both cases, the duration of the rectangular pulse is $30 \ \mu$s. 
The peak power is the same for all pulse shapes in (a), and the same for all pulse shapes in (b).
In (c), the total duration of each pulse is  $300 \ \mu$s and the peak power is varied to perfom a $\pi/2$ pulse.
In (d), the peak power is kept constant and the pulse duration is kept constant to $150 \ \mu$s.  
The maximum measured probabilities for the $\pi/2$ and $\pi$ pulse are  different from the ideal values of respectively $0.5$ and $1$ because of experimental imperfections (inhomogeneous Rabi frequency and imperfect normalization of the transition probability). 
}
\end{figure}

We investigate in this section the frequency selectivity of the  pulse shapes studied in this article, in line with previous works \cite{LuoYukun2016,Dunning2014}.
We measure the influence of the pulse shape on the frequency selectivity of the pulse, by varying the Raman laser frequency difference and measuring the transition probability. 
The results are presented in Fig.~\ref{fig:spectroscopy} (a) and (b) for a $\pi/2$ pulse and a $\pi$ pulse correspondingly, for the four pulse shapes investigated in the previous section: rectangular, Gaussian, GFlat, and GSinc.

The GSinc pulse is technically more difficult to implement than the other pulse shapes as it requires the introduction of phase jumps of $\pi$ at the points in time corresponding to the zeros of the power envelope in order to reverse the sign of the effective Rabi frequency (see Fig.~\ref{fig:time_trace_sinc} in the appendix for the time trace of the Sinc pulse).
The $\pi$ phase jumps are applied on the relative phase between the two Raman lasers through the phase lock loop, in a similar way as for the measurement of the sensitivity function presented in section \ref{subsec:measurement_g}.
For the data presented in panels (a)-(b), the GSinc is the product of a Gaussian  and of a Sinc function with 5 zeros on each side of the maximum (see appendix \ref{sec:pulsedef}).
The total duration of the pulse is $300 \ \mu$s, and the peak power is the same as for all pulse shapes.
The standard deviation of the Gaussian multiplying the Sinc function is $1/6$ of the total duration (i.e. $50 \ \mu$s).

\begin{figure}[!h]
 \includegraphics[width= \linewidth]{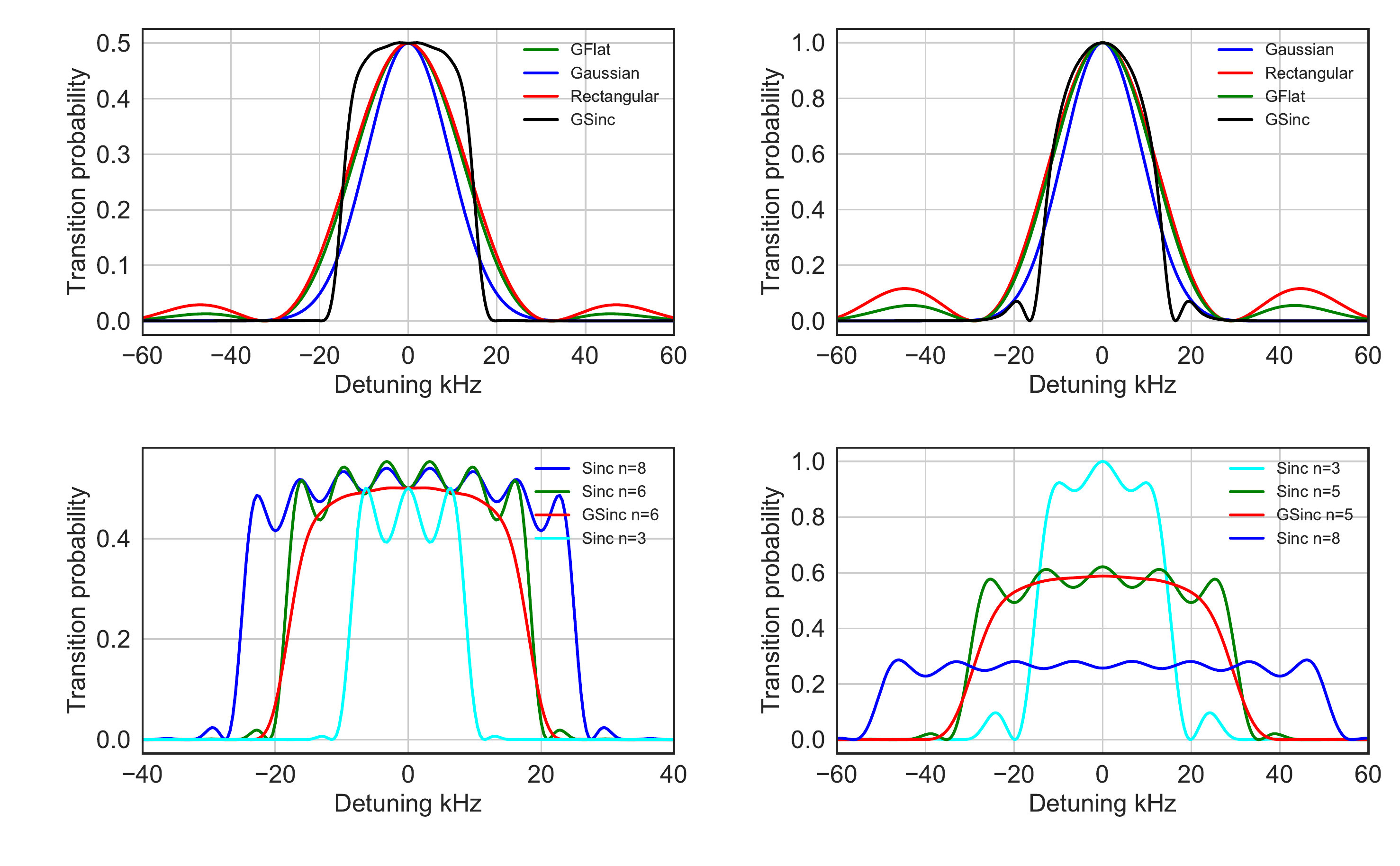}
\caption{\label{fig:spectroscopy_theory}
Calculations of the line shapes. The panels correspond to the measurement shown in Fig.~\ref{fig:spectroscopy}.   The parameters are fixed to the values measured in the experiment (pulse duration, peak Rabi frequency).
Note that the maximal probability of transition in this ideal calculation is $0.5$ for the $\pi / 2$ pulses and $1$ for the $\pi$ pulses.
}
\end{figure}
 
The experimental data are in agreement with the theoretical expectation, shown in Fig.~\ref{fig:spectroscopy_theory}, that the spectroscopy is the Fourier transform of the pulse shape.
In particular, the side lobes associated to the rectangular pulse are absent in the GFlat, Gaussian, and GSinc pulses.
The measurements also resolve the larger width of the GFlat pulse compared to the Gaussian pulse.
Finally, the GSinc pulse clearly shows sharper edges than the other pulse shapes.
The asymmetry in the GSinc spectroscopy is not fully understood: we think  that it is due to a non-linearity in the acousto-optic modulator which is driven for a longer duration for the GSinc pulse ($300 \ \mu$s) compared to the other pulse shapes (the spectroscopy were less asymetric when using shorter pulses).
   
We investigate experimentally in further details the influence of the number of zeros in the Sinc pulse on the sharpness of the spectroscopy.
The results are shown in Fig.~\ref{fig:spectroscopy} (c) and (d).
Panel (c) shows the measurements for a $\pi/2$ pulse where the total duration of all pulses is kept constant to $300 \ \mu$s, and the peak power is varied.
Panel (d) shows measurements were the peak power is kept constant and the pulse duration is kept constant to $150 \ \mu$s.

In conclusion, the Sinc and GSinc pulses exhibit an almost flat response to detuning, and a sharper decay than the other pulse shapes. They therefore optimizes the velocity acceptance of the pulse, at the expense of more complexity in the implementation.

\section{Scale factor of the interferometer}
\label{sec:scale_fact}

\begin{figure}[!bth]
\centering
\includegraphics[width = 0.7\linewidth]{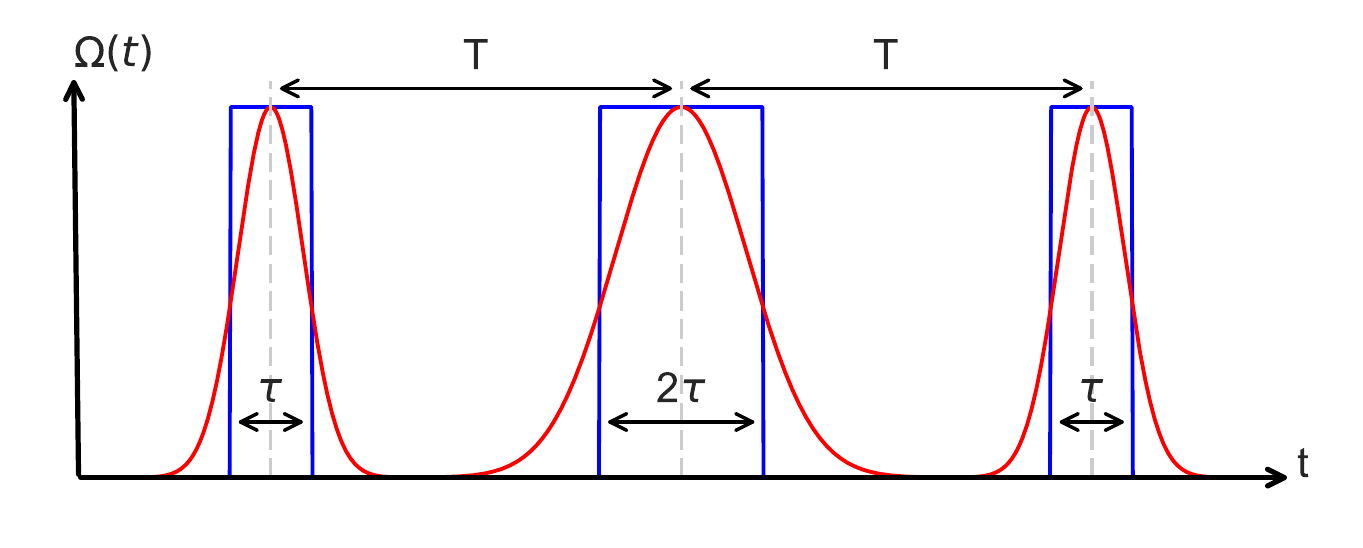}
\caption{\label{fig:T_definition} Illustration of the three-pulse
  interferometer sequence for rectangular and Gaussian pulse shapes.  The pulse separation $T$ denotes the time
  elapsed between the center of two consecutive light pulses, and $\tau$ is the duration of the $\pi/2$ rectangular pulse.
}
\end{figure}

The finite duration of the light pulses influences the scale factor of atom interferometers, i.e. their response to inertial effects.
The interferometer phase $\Phi$ is related to the relative laser phase $\phi(t)$ through the sensitivity function as $\Phi = \int g(t) \frac{d\phi}{dt} dt$.
Without loss of generality, we look at the example of a Mach-Zehnder-like interferometer sequence, where there are three light pulses ($\pi/2$-$\pi$-$\pi/2$ pulses) separated by $T$ between each consecutive pulse pairs.
See Fig.~\ref{fig:T_definition} for an illustration.
The finite duration $\tau$ of the $\pi/2$ (rectangular) pulses modifies the scale factor of an atom accelerometer from $\Phi = k_{\text{eff}} T^2 a$ to $\Phi=\mathcal{S}_{\text{rec}} a$, with $\mathcal{S}_{\text{rec}}=k_{\text{eff}} (T+\tau/2)\big(T+(\frac{4}{\pi}-\frac{3}{2})\tau\big)$ \cite{CheinetPhD}.
For experiments where the inertial effect is inferred from a phase measurement, such a change of scale factor has to be taken into account when evaluating the accuracy budget.

Furthermore, by modifying the temporal pulse shape, the scale factor $\mathcal{S}$ differs from that of rectangular pulses $\mathcal{S}_{\text{rec}}$.
Since $\tau/T$ is typically on the order of $10^{-4}$ or smaller, the relative correction $\frac{\mathcal{S}-\mathcal{S}_{\text{rec}}}{\mathcal{S}_{\text{rec}}}$ scales linearly with $\tau/T$, and can be evaluated numerically with the appropriate form of $g(t)$.
For example, for $T = 100$~ms and  a  peak Rabi frequency of $12.5$~kHz ($\tau = 10 \ \mu$s rectangular pulse), this correction amounts to $9.4\times 10^{-6}$ for a sequence of Gaussian pulses, $6.8\times 10^{-6}$ for GSinc pulses and $4.2\times 10^{-7}$ for GFlat pulses.

\section{Discussion}
\label{sec:discussion}

We summarize the properties of the four pulse shapes studied in this article in Table~\ref{tab:PulseComp}.
We report \textit{(i)} the velocity selectivity of a $\pi/2$ pulse (defined as the bandwith in units of the peak Rabi frequency, see appendix \ref{sec:pulsedef}), \textit{(ii)} the suitability for LMT interferometry, \textit{(iii)} the rejection of phase noise at high frequencies (according to section \ref{sec:noise}), and \textit{(iv)} the ease of implementation.
The main focus of this article was on the phase noise rejection.
Details on the velocity selectivity are given in appendix \ref{sec:pulsedef}.

Regarding LMT applications \cite{Muller2008a,Szigeti2012},  we extended the numerical calculations performed in \cite{RiouMielec2017} to implement arbitrary pulse shapes, and computed the Rabi oscillations for $10 \ \hbar k$ LMT  atom optics. 
We found that all smooth pulse shapes (Gaussian, GSinc, GFlat) support LMT beam splitters for pulse durations of few inverse peak Rabi frequency, in contrast to the rectangular pulse.

\begin{table*}[!h]
\centering
\begin{tabular}{|l||c|c|c|c|}
  \hline
  Pulse & Bandwidth (50\% $\mid$ 95\%) & LMT &  Noise rejection & Ease of implementation \\
  \hline
  Rectangular   & 1.73 $\mid$ 0.49  & Not suitable     & Weak, $1/f^2$ & Easiest \\
  Gaussian & 1.31 $\mid$  0.36  & Suitable & Strong & Medium \\
  GSinc    & 1.73 $\mid$ 1.01 & Suitable &  Strong & Difficult \\
  GFlat  & 1.65 $\mid$ 0.47  & Suitable & $1/f^6$ & Medium \\
  \hline
\end{tabular}
\caption{Summary of the properties of the  pulse shapes studied
  in this article. 
  The bandwith is defined as the two-photon detuning (in units of the peak Rabi frequency) where the transition probability falls to 50\% and 95\% of its maximum value.
   The phase noise   rejection (weak/strong) is defined according to the decay of the transfer function above the Rabi frequency, as shown in   Fig.~\ref{fig:Homega}.  \label{tab:PulseComp}}
\end{table*}

Regarding the ease of implementation, the rectangular
pulse is the most simple as it only requires a digital signal to
drive, typically, a voltage controlled oscillator. The implementation
of the Gaussian or the GFlat pulse shapes require a waveform generator and can be realized with relative ease.  The GSinc pulse (characterized by
negative values in the Rabi frequency) can be implemented
experimentally by setting $\pi$ phase shifts at the points of zero
crossing. It requires a waveform generator in combination
with a sufficiently fast phase modulation, and is thus more challenging to
implement.

Disregarding the implementation of the pulse shapes, the GSinc pulse is suited for all
applications, as it presents the largest velocity acceptance, can
efficiently perform LMT transitions, and rejects high frequency laser phase
noise. 
In comparison, although the GFlat pulse has a reduced velocity acceptance, it fulfills all other criteria, and can therefore be considered as a good compromise for various applications.

As a final note in this discussion, we remark that the interest of using an optical cavity to drive the light pulses in an AI has been raised recently \cite{Hamilton2015,RiouMielec2017}. The power enhancement at the cavity resonance requires sufficient finesse $\mathcal{F}$, which modifies the intensity build up time $\tau_{cav} = 2\mathcal{F}L/c$, and therefore the temporal shape of the pulse. The effect on the pulse shape will be particularly important in long-baseline gradiometers using AIs in an optical cavity, as planned in Ref.~\cite{Canuel_2014}, where $\tau_{cav}$ may be of the order of the pulse duration (i.e. few $\mu$s). We computed the sensitivity function for such a cavity-like pulse shape (see appendix \ref{sec:cavityTF}), which shows a $1/f^4$ high-frequency behavior.

\section{Conclusion}
\label{sec:conclusion}
We investigated the influence of temporally shaping the light pulses on the  response of an AI.
The main focus of our study was on the modification of the AI sensitivity function to phase, at frequencies of the order of and higher than the effective Rabi frequency. 
We demonstrated that smooth pulse shapes allow for a significant rejection of high frequency phase fluctuations  compared to rectangular pulses.
We also presented the modification of the scale factor of the AI due to pulse shaping, which has to be considered in the evaluation of systematic effects of AI sensors.
We finally discussed the trade-offs between the different representative pulse shapes considered in the article. 
One  important conclusion of our study is that  the rejection of high frequency phase fluctuations can be achieved with a minor effect on the velocity acceptance of the pulse by employing, for example, a GFlat pulse shape, which can also efficiently perform LMT beam splitters.

In the context of LMT interferometry, future work should study the modifications of the sensitivity function for AIs driven by LMT beam splitters (see Ref.~\cite{Decamps2016} for a prelimiary analysis) and  the influence on the rejection of the laser frequency noise, as has been done, for example, for laser intensity noise induced light shift~\cite{Clad_2010}.  

\section*{Acknowledgments}
\label{sec:Acknowledgments}
We acknowledge the financial support from Ville de Paris (project
HSENS-MWGRAV), FIRST-TF (ANR-10-LABX-48-01),   Centre National d'Etudes Saptiales (CNES), DIM
Nano-K, and Action Sp\'ecifique du CNRS Gravitation, R\'ef\'erences,
Astronomie et M\'etrologie (GRAM). B.F. is funded by Conseil
Scientifique de l'Observatoire de Paris, N.M. by Ville de Paris,
D.S. by Direction G\'en\'erale de l'Armement, and M.A. by the EDPIF doctoral school.  
We thank Azer Trim\`eche for his work on the programming of the arbitrary waveform generator used in this work,   Pierre
Dussarrat for pointing the GSinc pulse to our attention, and Leonid Sidorenkov for his contributions in the completion of this work.


\bibliography{PulseShaping}

\begin{thebibliography}{49}%
\makeatletter
\providecommand \@ifxundefined [1]{%
 \@ifx{#1\undefined}
}%
\providecommand \@ifnum [1]{%
 \ifnum #1\expandafter \@firstoftwo
 \else \expandafter \@secondoftwo
 \fi
}%
\providecommand \@ifx [1]{%
 \ifx #1\expandafter \@firstoftwo
 \else \expandafter \@secondoftwo
 \fi
}%
\providecommand \natexlab [1]{#1}%
\providecommand \enquote  [1]{``#1''}%
\providecommand \bibnamefont  [1]{#1}%
\providecommand \bibfnamefont [1]{#1}%
\providecommand \citenamefont [1]{#1}%
\providecommand \href@noop [0]{\@secondoftwo}%
\providecommand \href [0]{\begingroup \@sanitize@url \@href}%
\providecommand \@href[1]{\@@startlink{#1}\@@href}%
\providecommand \@@href[1]{\endgroup#1\@@endlink}%
\providecommand \@sanitize@url [0]{\catcode `\\12\catcode `\$12\catcode
  `\&12\catcode `\#12\catcode `\^12\catcode `\_12\catcode `\%12\relax}%
\providecommand \@@startlink[1]{}%
\providecommand \@@endlink[0]{}%
\providecommand \url  [0]{\begingroup\@sanitize@url \@url }%
\providecommand \@url [1]{\endgroup\@href {#1}{\urlprefix }}%
\providecommand \urlprefix  [0]{URL }%
\providecommand \Eprint [0]{\href }%
\providecommand \doibase [0]{http://dx.doi.org/}%
\providecommand \selectlanguage [0]{\@gobble}%
\providecommand \bibinfo  [0]{\@secondoftwo}%
\providecommand \bibfield  [0]{\@secondoftwo}%
\providecommand \translation [1]{[#1]}%
\providecommand \BibitemOpen [0]{}%
\providecommand \bibitemStop [0]{}%
\providecommand \bibitemNoStop [0]{.\EOS\space}%
\providecommand \EOS [0]{\spacefactor3000\relax}%
\providecommand \BibitemShut  [1]{\csname bibitem#1\endcsname}%
\let\auto@bib@innerbib\@empty
\bibitem [{\citenamefont {Canuel}\ \emph {et~al.}(2006)\citenamefont {Canuel},
  \citenamefont {Leduc}, \citenamefont {Holleville}, \citenamefont {Gauguet},
  \citenamefont {Fils}, \citenamefont {Virdis}, \citenamefont {Clairon},
  \citenamefont {Dimarcq}, \citenamefont {Bord{\'e}}, \citenamefont
  {Landragin},\ and\ \citenamefont {Bouyer}}]{Canuel2006}%
  \BibitemOpen
  \bibfield  {author} {\bibinfo {author} {\bibfnamefont {B.}~\bibnamefont
  {Canuel}}, \bibinfo {author} {\bibfnamefont {F.}~\bibnamefont {Leduc}},
  \bibinfo {author} {\bibfnamefont {D.}~\bibnamefont {Holleville}}, \bibinfo
  {author} {\bibfnamefont {A.}~\bibnamefont {Gauguet}}, \bibinfo {author}
  {\bibfnamefont {J.}~\bibnamefont {Fils}}, \bibinfo {author} {\bibfnamefont
  {A.}~\bibnamefont {Virdis}}, \bibinfo {author} {\bibfnamefont
  {A.}~\bibnamefont {Clairon}}, \bibinfo {author} {\bibfnamefont
  {N.}~\bibnamefont {Dimarcq}}, \bibinfo {author} {\bibfnamefont {C.~J.}\
  \bibnamefont {Bord{\'e}}}, \bibinfo {author} {\bibfnamefont {A.}~\bibnamefont
  {Landragin}}, \ and\ \bibinfo {author} {\bibfnamefont {P.}~\bibnamefont
  {Bouyer}},\ }\href {http://dx.doi.org/10.1103/PhysRevLett.97.010402}
  {\bibfield  {journal} {\bibinfo  {journal} {Phys.\ Rev.\ Lett.}\ }\textbf
  {\bibinfo {volume} {97}},\ \bibinfo {pages} {010402} (\bibinfo {year}
  {2006})}\BibitemShut {NoStop}%
\bibitem [{\citenamefont {Geiger}\ \emph {et~al.}(2011)\citenamefont {Geiger},
  \citenamefont {Menoret}, \citenamefont {Stern}, \citenamefont {Zahzam},
  \citenamefont {Cheinet}, \citenamefont {Battelier}, \citenamefont {Villing},
  \citenamefont {Moron}, \citenamefont {Lours}, \citenamefont {Bidel},
  \citenamefont {Bresson}, \citenamefont {Landragin},\ and\ \citenamefont
  {Bouyer}}]{Geiger2011}%
  \BibitemOpen
  \bibfield  {author} {\bibinfo {author} {\bibfnamefont {R.}~\bibnamefont
  {Geiger}}, \bibinfo {author} {\bibfnamefont {V.}~\bibnamefont {Menoret}},
  \bibinfo {author} {\bibfnamefont {G.}~\bibnamefont {Stern}}, \bibinfo
  {author} {\bibfnamefont {N.}~\bibnamefont {Zahzam}}, \bibinfo {author}
  {\bibfnamefont {P.}~\bibnamefont {Cheinet}}, \bibinfo {author} {\bibfnamefont
  {B.}~\bibnamefont {Battelier}}, \bibinfo {author} {\bibfnamefont
  {A.}~\bibnamefont {Villing}}, \bibinfo {author} {\bibfnamefont
  {F.}~\bibnamefont {Moron}}, \bibinfo {author} {\bibfnamefont
  {M.}~\bibnamefont {Lours}}, \bibinfo {author} {\bibfnamefont
  {Y.}~\bibnamefont {Bidel}}, \bibinfo {author} {\bibfnamefont
  {A.}~\bibnamefont {Bresson}}, \bibinfo {author} {\bibfnamefont
  {A.}~\bibnamefont {Landragin}}, \ and\ \bibinfo {author} {\bibfnamefont
  {P.}~\bibnamefont {Bouyer}},\ }\href {http://dx.doi.org/10.1038/ncomms1479}
  {\bibfield  {journal} {\bibinfo  {journal} {Nat. Commun.}\ }\textbf {\bibinfo
  {volume} {2}},\ \bibinfo {pages} {474} (\bibinfo {year} {2011})}\BibitemShut
  {NoStop}%
\bibitem [{\citenamefont {Rakholia}\ \emph {et~al.}(2014)\citenamefont
  {Rakholia}, \citenamefont {McGuinness},\ and\ \citenamefont
  {Biedermann}}]{Rakholia2014}%
  \BibitemOpen
  \bibfield  {author} {\bibinfo {author} {\bibfnamefont {A.~V.}\ \bibnamefont
  {Rakholia}}, \bibinfo {author} {\bibfnamefont {H.~J.}\ \bibnamefont
  {McGuinness}}, \ and\ \bibinfo {author} {\bibfnamefont {G.~W.}\ \bibnamefont
  {Biedermann}},\ }\href {\doibase 10.1103/physrevapplied.2.054012} {\bibfield
  {journal} {\bibinfo  {journal} {Phys. Rev. Appl.}\ }\textbf {\bibinfo
  {volume} {2}},\ \bibinfo {pages} {054012} (\bibinfo {year}
  {2014})}\BibitemShut {NoStop}%
\bibitem [{\citenamefont {Gustavson}\ \emph {et~al.}(2000)\citenamefont
  {Gustavson}, \citenamefont {Landragin},\ and\ \citenamefont
  {Kasevich}}]{Gustavson2000}%
  \BibitemOpen
  \bibfield  {author} {\bibinfo {author} {\bibfnamefont {T.~L.}\ \bibnamefont
  {Gustavson}}, \bibinfo {author} {\bibfnamefont {A.}~\bibnamefont
  {Landragin}}, \ and\ \bibinfo {author} {\bibfnamefont {M.~A.}\ \bibnamefont
  {Kasevich}},\ }\href {http://stacks.iop.org/0264-9381/17/i=12/a=311}
  {\bibfield  {journal} {\bibinfo  {journal} {Classical and Quantum Gravity}\
  }\textbf {\bibinfo {volume} {17}},\ \bibinfo {pages} {2385} (\bibinfo {year}
  {2000})}\BibitemShut {NoStop}%
\bibitem [{\citenamefont {Tackmann}\ \emph {et~al.}(2012)\citenamefont
  {Tackmann}, \citenamefont {Berg}, \citenamefont {Schubert}, \citenamefont
  {Abend}, \citenamefont {Gilowski}, \citenamefont {Ertmer},\ and\
  \citenamefont {Rasel}}]{Tackmann2012}%
  \BibitemOpen
  \bibfield  {author} {\bibinfo {author} {\bibfnamefont {G.}~\bibnamefont
  {Tackmann}}, \bibinfo {author} {\bibfnamefont {P.}~\bibnamefont {Berg}},
  \bibinfo {author} {\bibfnamefont {C.}~\bibnamefont {Schubert}}, \bibinfo
  {author} {\bibfnamefont {S.}~\bibnamefont {Abend}}, \bibinfo {author}
  {\bibfnamefont {M.}~\bibnamefont {Gilowski}}, \bibinfo {author}
  {\bibfnamefont {W.}~\bibnamefont {Ertmer}}, \ and\ \bibinfo {author}
  {\bibfnamefont {E.~M.}\ \bibnamefont {Rasel}},\ }\href
  {http://stacks.iop.org/1367-2630/14/i=1/a=015002} {\bibfield  {journal}
  {\bibinfo  {journal} {New Journal of Physics}\ }\textbf {\bibinfo {volume}
  {14}},\ \bibinfo {pages} {015002} (\bibinfo {year} {2012})}\BibitemShut
  {NoStop}%
\bibitem [{\citenamefont {Dutta}\ \emph {et~al.}(2016)\citenamefont {Dutta},
  \citenamefont {Savoie}, \citenamefont {Fang}, \citenamefont {Venon},
  \citenamefont {Alzar}, \citenamefont {Geiger},\ and\ \citenamefont
  {Landragin}}]{Dutta_2016}%
  \BibitemOpen
  \bibfield  {author} {\bibinfo {author} {\bibfnamefont {I.}~\bibnamefont
  {Dutta}}, \bibinfo {author} {\bibfnamefont {D.}~\bibnamefont {Savoie}},
  \bibinfo {author} {\bibfnamefont {B.}~\bibnamefont {Fang}}, \bibinfo {author}
  {\bibfnamefont {B.}~\bibnamefont {Venon}}, \bibinfo {author} {\bibfnamefont
  {C.~G.}\ \bibnamefont {Alzar}}, \bibinfo {author} {\bibfnamefont
  {R.}~\bibnamefont {Geiger}}, \ and\ \bibinfo {author} {\bibfnamefont
  {A.}~\bibnamefont {Landragin}},\ }\href {\doibase
  10.1103/physrevlett.116.183003} {\bibfield  {journal} {\bibinfo  {journal}
  {Physical Review Letters}\ }\textbf {\bibinfo {volume} {116}} (\bibinfo
  {year} {2016}),\ 10.1103/physrevlett.116.183003}\BibitemShut {NoStop}%
\bibitem [{\citenamefont {Louchet-Chauvet}\ \emph {et~al.}(2011)\citenamefont
  {Louchet-Chauvet}, \citenamefont {Farah}, \citenamefont {Bodart},
  \citenamefont {Clairon}, \citenamefont {Landragin}, \citenamefont {Merlet},\
  and\ \citenamefont {Pereira Dos~Santos}}]{LouchetChauvet2011}%
  \BibitemOpen
  \bibfield  {author} {\bibinfo {author} {\bibfnamefont {A.}~\bibnamefont
  {Louchet-Chauvet}}, \bibinfo {author} {\bibfnamefont {T.}~\bibnamefont
  {Farah}}, \bibinfo {author} {\bibfnamefont {Q.}~\bibnamefont {Bodart}},
  \bibinfo {author} {\bibfnamefont {A.}~\bibnamefont {Clairon}}, \bibinfo
  {author} {\bibfnamefont {A.}~\bibnamefont {Landragin}}, \bibinfo {author}
  {\bibfnamefont {S.}~\bibnamefont {Merlet}}, \ and\ \bibinfo {author}
  {\bibfnamefont {F.}~\bibnamefont {Pereira Dos~Santos}},\ }\href@noop {}
  {\bibfield  {journal} {\bibinfo  {journal} {New J. Phys.}\ }\textbf {\bibinfo
  {volume} {13}},\ \bibinfo {pages} {065025} (\bibinfo {year}
  {2011})}\BibitemShut {NoStop}%
\bibitem [{\citenamefont {Freier}\ \emph {et~al.}(2016)\citenamefont {Freier},
  \citenamefont {Hauth}, \citenamefont {Schkolnik}, \citenamefont {Leykauf},
  \citenamefont {Schilling}, \citenamefont {Wziontek}, \citenamefont
  {Scherneck}, \citenamefont {M\"{u}ller},\ and\ \citenamefont
  {Peters}}]{Freier2016}%
  \BibitemOpen
  \bibfield  {author} {\bibinfo {author} {\bibfnamefont {C.}~\bibnamefont
  {Freier}}, \bibinfo {author} {\bibfnamefont {M.}~\bibnamefont {Hauth}},
  \bibinfo {author} {\bibfnamefont {V.}~\bibnamefont {Schkolnik}}, \bibinfo
  {author} {\bibfnamefont {B.}~\bibnamefont {Leykauf}}, \bibinfo {author}
  {\bibfnamefont {M.}~\bibnamefont {Schilling}}, \bibinfo {author}
  {\bibfnamefont {H.}~\bibnamefont {Wziontek}}, \bibinfo {author}
  {\bibfnamefont {H.-G.}\ \bibnamefont {Scherneck}}, \bibinfo {author}
  {\bibfnamefont {J.}~\bibnamefont {M\"{u}ller}}, \ and\ \bibinfo {author}
  {\bibfnamefont {A.}~\bibnamefont {Peters}},\ }\href {\doibase
  10.1088/1742-6596/723/1/012050} {\bibfield  {journal} {\bibinfo  {journal}
  {J. Phys. Conf. Ser.}\ }\textbf {\bibinfo {volume} {723}},\ \bibinfo {pages}
  {012050} (\bibinfo {year} {2016})}\BibitemShut {NoStop}%
\bibitem [{\citenamefont {McGuirk}\ \emph {et~al.}(2002)\citenamefont
  {McGuirk}, \citenamefont {Foster}, \citenamefont {Fixler}, \citenamefont
  {Snadden},\ and\ \citenamefont {Kasevich}}]{McGuirk2002}%
  \BibitemOpen
  \bibfield  {author} {\bibinfo {author} {\bibfnamefont {J.~M.}\ \bibnamefont
  {McGuirk}}, \bibinfo {author} {\bibfnamefont {G.~T.}\ \bibnamefont {Foster}},
  \bibinfo {author} {\bibfnamefont {J.~B.}\ \bibnamefont {Fixler}}, \bibinfo
  {author} {\bibfnamefont {M.~J.}\ \bibnamefont {Snadden}}, \ and\ \bibinfo
  {author} {\bibfnamefont {M.~A.}\ \bibnamefont {Kasevich}},\ }\href {\doibase
  10.1103/PhysRevA.65.033608} {\bibfield  {journal} {\bibinfo  {journal} {Phys.
  Rev. A}\ }\textbf {\bibinfo {volume} {65}},\ \bibinfo {pages} {033608}
  (\bibinfo {year} {2002})}\BibitemShut {NoStop}%
\bibitem [{\citenamefont {Rosi}\ \emph {et~al.}(2015)\citenamefont {Rosi},
  \citenamefont {Cacciapuoti}, \citenamefont {Sorrentino}, \citenamefont
  {Menchetti}, \citenamefont {Prevedelli},\ and\ \citenamefont
  {Tino}}]{Rosi2015}%
  \BibitemOpen
  \bibfield  {author} {\bibinfo {author} {\bibfnamefont {G.}~\bibnamefont
  {Rosi}}, \bibinfo {author} {\bibfnamefont {L.}~\bibnamefont {Cacciapuoti}},
  \bibinfo {author} {\bibfnamefont {F.}~\bibnamefont {Sorrentino}}, \bibinfo
  {author} {\bibfnamefont {M.}~\bibnamefont {Menchetti}}, \bibinfo {author}
  {\bibfnamefont {M.}~\bibnamefont {Prevedelli}}, \ and\ \bibinfo {author}
  {\bibfnamefont {G.~M.}\ \bibnamefont {Tino}},\ }\href
  {http://dx.doi.org/10.1103/PhysRevLett.114.013001} {\bibfield  {journal}
  {\bibinfo  {journal} {Phys.\ Rev.\ Lett.}\ }\textbf {\bibinfo {volume}
  {114}},\ \bibinfo {pages} {013001} (\bibinfo {year} {2015})}\BibitemShut
  {NoStop}%
\bibitem [{\citenamefont {Bouchendira}\ \emph {et~al.}(2011)\citenamefont
  {Bouchendira}, \citenamefont {Clad{\'e}}, \citenamefont
  {Guellati-Kh{\'e}lifa}, \citenamefont {Nez},\ and\ \citenamefont
  {Biraben}}]{Bouchendira2011}%
  \BibitemOpen
  \bibfield  {author} {\bibinfo {author} {\bibfnamefont {R.}~\bibnamefont
  {Bouchendira}}, \bibinfo {author} {\bibfnamefont {P.}~\bibnamefont
  {Clad{\'e}}}, \bibinfo {author} {\bibfnamefont {S.}~\bibnamefont
  {Guellati-Kh{\'e}lifa}}, \bibinfo {author} {\bibfnamefont {F.}~\bibnamefont
  {Nez}}, \ and\ \bibinfo {author} {\bibfnamefont {F.}~\bibnamefont
  {Biraben}},\ }\href {http://dx.doi.org/10.1103/PhysRevLett.106.080801}
  {\bibfield  {journal} {\bibinfo  {journal} {Phys.\ Rev.\ Lett.}\ }\textbf
  {\bibinfo {volume} {106}},\ \bibinfo {pages} {080801} (\bibinfo {year}
  {2011})}\BibitemShut {NoStop}%
\bibitem [{\citenamefont {Rosi}\ \emph {et~al.}(2014)\citenamefont {Rosi},
  \citenamefont {Sorrentino}, \citenamefont {Cacciapuoti}, \citenamefont
  {Prevedelli},\ and\ \citenamefont {Tino}}]{Rosi2014}%
  \BibitemOpen
  \bibfield  {author} {\bibinfo {author} {\bibfnamefont {G.}~\bibnamefont
  {Rosi}}, \bibinfo {author} {\bibfnamefont {F.}~\bibnamefont {Sorrentino}},
  \bibinfo {author} {\bibfnamefont {L.}~\bibnamefont {Cacciapuoti}}, \bibinfo
  {author} {\bibfnamefont {M.}~\bibnamefont {Prevedelli}}, \ and\ \bibinfo
  {author} {\bibfnamefont {G.}~\bibnamefont {Tino}},\ }\href
  {http://dx.doi.org/10.1038/nature13433} {\bibfield  {journal} {\bibinfo
  {journal} {Nature}\ }\textbf {\bibinfo {volume} {510}},\ \bibinfo {pages}
  {518} (\bibinfo {year} {2014})}\BibitemShut {NoStop}%
\bibitem [{\citenamefont {Varoquaux}\ \emph {et~al.}(2009)\citenamefont
  {Varoquaux}, \citenamefont {Nyman}, \citenamefont {Geiger}, \citenamefont
  {Cheinet}, \citenamefont {Landragin},\ and\ \citenamefont
  {Bouyer}}]{Varoquaux2009}%
  \BibitemOpen
  \bibfield  {author} {\bibinfo {author} {\bibfnamefont {G.}~\bibnamefont
  {Varoquaux}}, \bibinfo {author} {\bibfnamefont {R.~A.}\ \bibnamefont
  {Nyman}}, \bibinfo {author} {\bibfnamefont {R.}~\bibnamefont {Geiger}},
  \bibinfo {author} {\bibfnamefont {P.}~\bibnamefont {Cheinet}}, \bibinfo
  {author} {\bibfnamefont {A.}~\bibnamefont {Landragin}}, \ and\ \bibinfo
  {author} {\bibfnamefont {P.}~\bibnamefont {Bouyer}},\ }\href
  {http://stacks.iop.org/1367-2630/11/i=11/a=113010} {\bibfield  {journal}
  {\bibinfo  {journal} {New Journal of Physics}\ }\textbf {\bibinfo {volume}
  {11}},\ \bibinfo {pages} {113010} (\bibinfo {year} {2009})}\BibitemShut
  {NoStop}%
\bibitem [{\citenamefont {Schlippert}\ \emph {et~al.}(2014)\citenamefont
  {Schlippert}, \citenamefont {Hartwig}, \citenamefont {Albers}, \citenamefont
  {Richardson}, \citenamefont {Schubert}, \citenamefont {Roura}, \citenamefont
  {Schleich}, \citenamefont {Ertmer},\ and\ \citenamefont
  {Rasel}}]{Schlippert2014}%
  \BibitemOpen
  \bibfield  {author} {\bibinfo {author} {\bibfnamefont {D.}~\bibnamefont
  {Schlippert}}, \bibinfo {author} {\bibfnamefont {J.}~\bibnamefont {Hartwig}},
  \bibinfo {author} {\bibfnamefont {H.}~\bibnamefont {Albers}}, \bibinfo
  {author} {\bibfnamefont {L.~L.}\ \bibnamefont {Richardson}}, \bibinfo
  {author} {\bibfnamefont {C.}~\bibnamefont {Schubert}}, \bibinfo {author}
  {\bibfnamefont {A.}~\bibnamefont {Roura}}, \bibinfo {author} {\bibfnamefont
  {W.~P.}\ \bibnamefont {Schleich}}, \bibinfo {author} {\bibfnamefont
  {W.}~\bibnamefont {Ertmer}}, \ and\ \bibinfo {author} {\bibfnamefont {E.~M.}\
  \bibnamefont {Rasel}},\ }\href {\doibase 10.1103/PhysRevLett.112.203002}
  {\bibfield  {journal} {\bibinfo  {journal} {Phys. Rev. Lett.}\ }\textbf
  {\bibinfo {volume} {112}},\ \bibinfo {pages} {203002} (\bibinfo {year}
  {2014})}\BibitemShut {NoStop}%
\bibitem [{\citenamefont {Zhou}\ \emph {et~al.}(2015)\citenamefont {Zhou},
  \citenamefont {Long}, \citenamefont {Tang}, \citenamefont {Chen},
  \citenamefont {Gao}, \citenamefont {Peng}, \citenamefont {Duan},
  \citenamefont {Zhong}, \citenamefont {Xiong}, \citenamefont {Wang},
  \citenamefont {Zhang},\ and\ \citenamefont {Zhan}}]{Zhou2015}%
  \BibitemOpen
  \bibfield  {author} {\bibinfo {author} {\bibfnamefont {L.}~\bibnamefont
  {Zhou}}, \bibinfo {author} {\bibfnamefont {S.}~\bibnamefont {Long}}, \bibinfo
  {author} {\bibfnamefont {B.}~\bibnamefont {Tang}}, \bibinfo {author}
  {\bibfnamefont {X.}~\bibnamefont {Chen}}, \bibinfo {author} {\bibfnamefont
  {F.}~\bibnamefont {Gao}}, \bibinfo {author} {\bibfnamefont {W.}~\bibnamefont
  {Peng}}, \bibinfo {author} {\bibfnamefont {W.}~\bibnamefont {Duan}}, \bibinfo
  {author} {\bibfnamefont {J.}~\bibnamefont {Zhong}}, \bibinfo {author}
  {\bibfnamefont {Z.}~\bibnamefont {Xiong}}, \bibinfo {author} {\bibfnamefont
  {J.}~\bibnamefont {Wang}}, \bibinfo {author} {\bibfnamefont {Y.}~\bibnamefont
  {Zhang}}, \ and\ \bibinfo {author} {\bibfnamefont {M.}~\bibnamefont {Zhan}},\
  }\href {\doibase 10.1103/PhysRevLett.115.013004} {\bibfield  {journal}
  {\bibinfo  {journal} {Phys. Rev. Lett.}\ }\textbf {\bibinfo {volume} {115}},\
  \bibinfo {pages} {013004} (\bibinfo {year} {2015})}\BibitemShut {NoStop}%
\bibitem [{\citenamefont {Bonnin}\ \emph {et~al.}(2013)\citenamefont {Bonnin},
  \citenamefont {Zahzam}, \citenamefont {Y}, \citenamefont {Bidel},\ and\
  \citenamefont {Bresson}}]{Bonnin2013}%
  \BibitemOpen
  \bibfield  {author} {\bibinfo {author} {\bibfnamefont {A.}~\bibnamefont
  {Bonnin}}, \bibinfo {author} {\bibfnamefont {N.}~\bibnamefont {Zahzam}},
  \bibinfo {author} {\bibnamefont {Y}}, \bibinfo {author} {\bibnamefont
  {Bidel}}, \ and\ \bibinfo {author} {\bibfnamefont {A.}~\bibnamefont
  {Bresson}},\ }\href@noop {} {\bibfield  {journal} {\bibinfo  {journal} {Phys.
  Rev. A}\ }\textbf {\bibinfo {volume} {88}},\ \bibinfo {pages} {043615}
  (\bibinfo {year} {2013})}\BibitemShut {NoStop}%
\bibitem [{\citenamefont {Duan}\ \emph {et~al.}(2016)\citenamefont {Duan},
  \citenamefont {Deng}, \citenamefont {Zhou}, \citenamefont {Zhang},
  \citenamefont {Xu}, \citenamefont {Xiong}, \citenamefont {Xu}, \citenamefont
  {Shao}, \citenamefont {Luo},\ and\ \citenamefont {Hu}}]{Duan2016}%
  \BibitemOpen
  \bibfield  {author} {\bibinfo {author} {\bibfnamefont {X.-C.}\ \bibnamefont
  {Duan}}, \bibinfo {author} {\bibfnamefont {X.-B.}\ \bibnamefont {Deng}},
  \bibinfo {author} {\bibfnamefont {M.-K.}\ \bibnamefont {Zhou}}, \bibinfo
  {author} {\bibfnamefont {K.}~\bibnamefont {Zhang}}, \bibinfo {author}
  {\bibfnamefont {W.-J.}\ \bibnamefont {Xu}}, \bibinfo {author} {\bibfnamefont
  {F.}~\bibnamefont {Xiong}}, \bibinfo {author} {\bibfnamefont {Y.-Y.}\
  \bibnamefont {Xu}}, \bibinfo {author} {\bibfnamefont {C.-G.}\ \bibnamefont
  {Shao}}, \bibinfo {author} {\bibfnamefont {J.}~\bibnamefont {Luo}}, \ and\
  \bibinfo {author} {\bibfnamefont {Z.-K.}\ \bibnamefont {Hu}},\ }\href
  {\doibase 10.1103/PhysRevLett.117.023001} {\bibfield  {journal} {\bibinfo
  {journal} {Phys. Rev. Lett.}\ }\textbf {\bibinfo {volume} {117}},\ \bibinfo
  {pages} {023001} (\bibinfo {year} {2016})}\BibitemShut {NoStop}%
\bibitem [{\citenamefont {{D. Aguilera et al}}(2014)}]{Aguilera2014}%
  \BibitemOpen
  \bibfield  {author} {\bibinfo {author} {\bibnamefont {{D. Aguilera et al}}},\
  }\href
  {http://iopscience.iop.org/article/10.1088/0264-9381/31/11/115010/meta}
  {\bibfield  {journal} {\bibinfo  {journal} {Classical Quantum Gravity}\
  }\textbf {\bibinfo {volume} {31}},\ \bibinfo {pages} {115010} (\bibinfo
  {year} {2014})}\BibitemShut {NoStop}%
\bibitem [{\citenamefont {Barrett}\ \emph {et~al.}(2016)\citenamefont
  {Barrett}, \citenamefont {Antoni-Micollier}, \citenamefont {Chichet},
  \citenamefont {Battelier}, \citenamefont {Lévèque}, \citenamefont
  {Landragin},\ and\ \citenamefont {Bouyer}}]{Barrett2016}%
  \BibitemOpen
  \bibfield  {author} {\bibinfo {author} {\bibfnamefont {B.}~\bibnamefont
  {Barrett}}, \bibinfo {author} {\bibfnamefont {L.}~\bibnamefont
  {Antoni-Micollier}}, \bibinfo {author} {\bibfnamefont {L.}~\bibnamefont
  {Chichet}}, \bibinfo {author} {\bibfnamefont {B.}~\bibnamefont {Battelier}},
  \bibinfo {author} {\bibfnamefont {T.}~\bibnamefont {Lévèque}}, \bibinfo
  {author} {\bibfnamefont {A.}~\bibnamefont {Landragin}}, \ and\ \bibinfo
  {author} {\bibfnamefont {P.}~\bibnamefont {Bouyer}},\ }\href
  {http://dx.doi.org/10.1038/ncomms13786} {\bibfield  {journal} {\bibinfo
  {journal} {Nature Communications}\ }\textbf {\bibinfo {volume} {7}},\
  \bibinfo {pages} {13786} (\bibinfo {year} {2016})}\BibitemShut {NoStop}%
\bibitem [{\citenamefont {Rosi}\ \emph {et~al.}(2017)\citenamefont {Rosi},
  \citenamefont {D'Amico}, \citenamefont {Cacciapuoti}, \citenamefont
  {Sorrentino}, \citenamefont {Prevedelli}, \citenamefont {Zych}, \citenamefont
  {Brukner},\ and\ \citenamefont {Tino}}]{Rosi2017}%
  \BibitemOpen
  \bibfield  {author} {\bibinfo {author} {\bibfnamefont {G.}~\bibnamefont
  {Rosi}}, \bibinfo {author} {\bibfnamefont {G.}~\bibnamefont {D'Amico}},
  \bibinfo {author} {\bibfnamefont {L.}~\bibnamefont {Cacciapuoti}}, \bibinfo
  {author} {\bibfnamefont {F.}~\bibnamefont {Sorrentino}}, \bibinfo {author}
  {\bibfnamefont {M.}~\bibnamefont {Prevedelli}}, \bibinfo {author}
  {\bibfnamefont {M.}~\bibnamefont {Zych}}, \bibinfo {author} {\bibfnamefont
  {C.}~\bibnamefont {Brukner}}, \ and\ \bibinfo {author} {\bibfnamefont
  {G.~M.}\ \bibnamefont {Tino}},\ }\href
  {http://dx.doi.org/10.1038/ncomms15529} {\bibfield  {journal} {\bibinfo
  {journal} {Nat. Commun.}\ }\textbf {\bibinfo {volume} {8}},\ \bibinfo {pages}
  {15529} (\bibinfo {year} {2017})}\BibitemShut {NoStop}%
\bibitem [{\citenamefont {{Overstreet}}\ \emph {et~al.}(2017)\citenamefont
  {{Overstreet}}, \citenamefont {{Asenbaum}}, \citenamefont {{Kovachy}},
  \citenamefont {{Notermans}}, \citenamefont {{Hogan}},\ and\ \citenamefont
  {{Kasevich}}}]{Overstreet2017}%
  \BibitemOpen
  \bibfield  {author} {\bibinfo {author} {\bibfnamefont {C.}~\bibnamefont
  {{Overstreet}}}, \bibinfo {author} {\bibfnamefont {P.}~\bibnamefont
  {{Asenbaum}}}, \bibinfo {author} {\bibfnamefont {T.}~\bibnamefont
  {{Kovachy}}}, \bibinfo {author} {\bibfnamefont {R.}~\bibnamefont
  {{Notermans}}}, \bibinfo {author} {\bibfnamefont {J.~M.}\ \bibnamefont
  {{Hogan}}}, \ and\ \bibinfo {author} {\bibfnamefont {M.~A.}\ \bibnamefont
  {{Kasevich}}},\ }\href@noop {} {\bibfield  {journal} {\bibinfo  {journal}
  {ArXiv e-prints}\ } (\bibinfo {year} {2017})},\ \Eprint
  {http://arxiv.org/abs/1711.09986} {arXiv:1711.09986 [physics.atom-ph]}
  \BibitemShut {NoStop}%
\bibitem [{\citenamefont {Dimopoulos}\ \emph {et~al.}(2009)\citenamefont
  {Dimopoulos}, \citenamefont {Graham}, \citenamefont {Hogan}, \citenamefont
  {Kasevich},\ and\ \citenamefont {Rajendran}}]{Dimopoulos2009}%
  \BibitemOpen
  \bibfield  {author} {\bibinfo {author} {\bibfnamefont {S.}~\bibnamefont
  {Dimopoulos}}, \bibinfo {author} {\bibfnamefont {P.~W.}\ \bibnamefont
  {Graham}}, \bibinfo {author} {\bibfnamefont {J.~M.}\ \bibnamefont {Hogan}},
  \bibinfo {author} {\bibfnamefont {M.~A.}\ \bibnamefont {Kasevich}}, \ and\
  \bibinfo {author} {\bibfnamefont {S.}~\bibnamefont {Rajendran}},\ }\href
  {http://www.sciencedirect.com/science/article/pii/S0370269309006844}
  {\bibfield  {journal} {\bibinfo  {journal} {Physics Letters B}\ }\textbf
  {\bibinfo {volume} {678}},\ \bibinfo {pages} {37} (\bibinfo {year}
  {2009})}\BibitemShut {NoStop}%
\bibitem [{\citenamefont {Geiger}(2016)}]{Geiger2016}%
  \BibitemOpen
  \bibfield  {author} {\bibinfo {author} {\bibfnamefont {R.}~\bibnamefont
  {Geiger}},\ }in\ \href {\doibase 10.1142/9789813141766_0008} {\emph {\bibinfo
  {booktitle} {Future Gravitational Wave Detectors Based on Atom
  Interferometry,}}}\ (\bibinfo  {publisher} {\textsl{An Overview of
  Gravitational Waves : Theory, Sources and Detection}, edited by Gerard Auger
  and Eric Plagnol, World Scientific},\ \bibinfo {year} {2016})\ pp.\ \bibinfo
  {pages} {285--313}\BibitemShut {NoStop}%
\bibitem [{\citenamefont {Hogan}\ \emph {et~al.}(2011)\citenamefont {Hogan},
  \citenamefont {Johnson}, \citenamefont {Dickerson}, \citenamefont {Kovachy},
  \citenamefont {Sugarbaker}, \citenamefont {Chiow}, \citenamefont {Graham},
  \citenamefont {Kasevich}, \citenamefont {Saif}, \citenamefont {Rajendran},
  \citenamefont {Bouyer}, \citenamefont {Seery}, \citenamefont {Feinberg},\
  and\ \citenamefont {Keski-Kuha}}]{Hogan2011}%
  \BibitemOpen
  \bibfield  {author} {\bibinfo {author} {\bibfnamefont {J.~M.}\ \bibnamefont
  {Hogan}}, \bibinfo {author} {\bibfnamefont {D.~M.~S.}\ \bibnamefont
  {Johnson}}, \bibinfo {author} {\bibfnamefont {S.}~\bibnamefont {Dickerson}},
  \bibinfo {author} {\bibfnamefont {T.}~\bibnamefont {Kovachy}}, \bibinfo
  {author} {\bibfnamefont {A.}~\bibnamefont {Sugarbaker}}, \bibinfo {author}
  {\bibfnamefont {S.-w.}\ \bibnamefont {Chiow}}, \bibinfo {author}
  {\bibfnamefont {P.~W.}\ \bibnamefont {Graham}}, \bibinfo {author}
  {\bibfnamefont {M.~A.}\ \bibnamefont {Kasevich}}, \bibinfo {author}
  {\bibfnamefont {B.}~\bibnamefont {Saif}}, \bibinfo {author} {\bibfnamefont
  {S.}~\bibnamefont {Rajendran}}, \bibinfo {author} {\bibfnamefont
  {P.}~\bibnamefont {Bouyer}}, \bibinfo {author} {\bibfnamefont {B.~D.}\
  \bibnamefont {Seery}}, \bibinfo {author} {\bibfnamefont {L.}~\bibnamefont
  {Feinberg}}, \ and\ \bibinfo {author} {\bibfnamefont {R.}~\bibnamefont
  {Keski-Kuha}},\ }\href {http://dx.doi.org/10.1007/s10714-011-1182-x}
  {\bibfield  {journal} {\bibinfo  {journal} {General Relativity and
  Gravitation}\ }\textbf {\bibinfo {volume} {43}},\ \bibinfo {pages} {1953}
  (\bibinfo {year} {2011})}\BibitemShut {NoStop}%
\bibitem [{\citenamefont {Kasevich}\ \emph {et~al.}(1991)\citenamefont
  {Kasevich}, \citenamefont {Weiss}, \citenamefont {Riis}, \citenamefont
  {Moler}, \citenamefont {Kasapi},\ and\ \citenamefont
  {Chu}}]{Kasevich1991vel}%
  \BibitemOpen
  \bibfield  {author} {\bibinfo {author} {\bibfnamefont {M.}~\bibnamefont
  {Kasevich}}, \bibinfo {author} {\bibfnamefont {D.~S.}\ \bibnamefont {Weiss}},
  \bibinfo {author} {\bibfnamefont {E.}~\bibnamefont {Riis}}, \bibinfo {author}
  {\bibfnamefont {K.}~\bibnamefont {Moler}}, \bibinfo {author} {\bibfnamefont
  {S.}~\bibnamefont {Kasapi}}, \ and\ \bibinfo {author} {\bibfnamefont
  {S.}~\bibnamefont {Chu}},\ }\href {\doibase 10.1103/PhysRevLett.66.2297}
  {\bibfield  {journal} {\bibinfo  {journal} {Phys. Rev. Lett.}\ }\textbf
  {\bibinfo {volume} {66}},\ \bibinfo {pages} {2297} (\bibinfo {year}
  {1991})}\BibitemShut {NoStop}%
\bibitem [{\citenamefont {Luo}\ \emph {et~al.}(2016)\citenamefont {Luo},
  \citenamefont {Yan}, \citenamefont {Hu}, \citenamefont {Jia}, \citenamefont
  {Wei},\ and\ \citenamefont {Yang}}]{LuoYukun2016}%
  \BibitemOpen
  \bibfield  {author} {\bibinfo {author} {\bibfnamefont {Y.}~\bibnamefont
  {Luo}}, \bibinfo {author} {\bibfnamefont {S.}~\bibnamefont {Yan}}, \bibinfo
  {author} {\bibfnamefont {Q.}~\bibnamefont {Hu}}, \bibinfo {author}
  {\bibfnamefont {A.}~\bibnamefont {Jia}}, \bibinfo {author} {\bibfnamefont
  {C.}~\bibnamefont {Wei}}, \ and\ \bibinfo {author} {\bibfnamefont
  {J.}~\bibnamefont {Yang}},\ }\href {\doibase 10.1140/epjd/e2016-70428-6}
  {\bibfield  {journal} {\bibinfo  {journal} {Eur. Phys. J. D}\ }\textbf
  {\bibinfo {volume} {70}},\ \bibinfo {pages} {262} (\bibinfo {year}
  {2016})}\BibitemShut {NoStop}%
\bibitem [{\citenamefont {Dunning}\ \emph {et~al.}(2014)\citenamefont
  {Dunning}, \citenamefont {Gregory}, \citenamefont {Bateman}, \citenamefont
  {Cooper}, \citenamefont {Himsworth}, \citenamefont {Jones},\ and\
  \citenamefont {Freegarde}}]{Dunning2014}%
  \BibitemOpen
  \bibfield  {author} {\bibinfo {author} {\bibfnamefont {A.}~\bibnamefont
  {Dunning}}, \bibinfo {author} {\bibfnamefont {R.}~\bibnamefont {Gregory}},
  \bibinfo {author} {\bibfnamefont {J.}~\bibnamefont {Bateman}}, \bibinfo
  {author} {\bibfnamefont {N.}~\bibnamefont {Cooper}}, \bibinfo {author}
  {\bibfnamefont {M.}~\bibnamefont {Himsworth}}, \bibinfo {author}
  {\bibfnamefont {J.~A.}\ \bibnamefont {Jones}}, \ and\ \bibinfo {author}
  {\bibfnamefont {T.}~\bibnamefont {Freegarde}},\ }\href {\doibase
  10.1103/physreva.90.033608} {\bibfield  {journal} {\bibinfo  {journal} {Phys.
  Rev. A}\ }\textbf {\bibinfo {volume} {90}} (\bibinfo {year} {2014}),\
  10.1103/physreva.90.033608}\BibitemShut {NoStop}%
\bibitem [{\citenamefont {M\"{u}ller}\ \emph {et~al.}(2008)\citenamefont
  {M\"{u}ller}, \citenamefont {Chiow},\ and\ \citenamefont
  {Chu}}]{Muller2008a}%
  \BibitemOpen
  \bibfield  {author} {\bibinfo {author} {\bibfnamefont {H.}~\bibnamefont
  {M\"{u}ller}}, \bibinfo {author} {\bibfnamefont {S.}~\bibnamefont {Chiow}}, \
  and\ \bibinfo {author} {\bibfnamefont {S.}~\bibnamefont {Chu}},\ }\href
  {\doibase 10.1103/physreva.77.023609} {\bibfield  {journal} {\bibinfo
  {journal} {Phys. Rev. A}\ }\textbf {\bibinfo {volume} {77}},\ \bibinfo
  {pages} {023609} (\bibinfo {year} {2008})}\BibitemShut {NoStop}%
\bibitem [{\citenamefont {Szigeti}\ \emph {et~al.}(2012)\citenamefont
  {Szigeti}, \citenamefont {Debs}, \citenamefont {Hope}, \citenamefont
  {Robins},\ and\ \citenamefont {Close}}]{Szigeti2012}%
  \BibitemOpen
  \bibfield  {author} {\bibinfo {author} {\bibfnamefont {S.~S.}\ \bibnamefont
  {Szigeti}}, \bibinfo {author} {\bibfnamefont {J.~E.}\ \bibnamefont {Debs}},
  \bibinfo {author} {\bibfnamefont {J.~J.}\ \bibnamefont {Hope}}, \bibinfo
  {author} {\bibfnamefont {N.~P.}\ \bibnamefont {Robins}}, \ and\ \bibinfo
  {author} {\bibfnamefont {J.~D.}\ \bibnamefont {Close}},\ }\href
  {http://stacks.iop.org/1367-2630/14/i=2/a=023009} {\bibfield  {journal}
  {\bibinfo  {journal} {New Journal of Physics}\ }\textbf {\bibinfo {volume}
  {14}},\ \bibinfo {pages} {023009} (\bibinfo {year} {2012})}\BibitemShut
  {NoStop}%
\bibitem [{\citenamefont {Santos}\ \emph {et~al.}(2002)\citenamefont {Santos},
  \citenamefont {Marion}, \citenamefont {Bize}, \citenamefont {Sortais},
  \citenamefont {Clairon},\ and\ \citenamefont
  {Salomon}}]{Pereira_Dos_Santos_2002}%
  \BibitemOpen
  \bibfield  {author} {\bibinfo {author} {\bibfnamefont {F.~P.~D.}\
  \bibnamefont {Santos}}, \bibinfo {author} {\bibfnamefont {H.}~\bibnamefont
  {Marion}}, \bibinfo {author} {\bibfnamefont {S.}~\bibnamefont {Bize}},
  \bibinfo {author} {\bibfnamefont {Y.}~\bibnamefont {Sortais}}, \bibinfo
  {author} {\bibfnamefont {A.}~\bibnamefont {Clairon}}, \ and\ \bibinfo
  {author} {\bibfnamefont {C.}~\bibnamefont {Salomon}},\ }\href {\doibase
  10.1103/physrevlett.89.233004} {\bibfield  {journal} {\bibinfo  {journal}
  {Physical Review Letters}\ }\textbf {\bibinfo {volume} {89}},\ \bibinfo
  {pages} {233004} (\bibinfo {year} {2002})}\BibitemShut {NoStop}%
\bibitem [{\citenamefont {Kovachy}\ \emph {et~al.}(2012)\citenamefont
  {Kovachy}, \citenamefont {wey Chiow},\ and\ \citenamefont
  {Kasevich}}]{Kovachy_2012}%
  \BibitemOpen
  \bibfield  {author} {\bibinfo {author} {\bibfnamefont {T.}~\bibnamefont
  {Kovachy}}, \bibinfo {author} {\bibfnamefont {S.}~\bibnamefont {wey Chiow}},
  \ and\ \bibinfo {author} {\bibfnamefont {M.~A.}\ \bibnamefont {Kasevich}},\
  }\href {\doibase 10.1103/physreva.86.011606} {\bibfield  {journal} {\bibinfo
  {journal} {Physical Review A}\ }\textbf {\bibinfo {volume} {86}},\ \bibinfo
  {pages} {011606(R)} (\bibinfo {year} {2012})}\BibitemShut {NoStop}%
\bibitem [{\citenamefont {Dick}(1987)}]{Dick_1987}%
  \BibitemOpen
  \bibfield  {author} {\bibinfo {author} {\bibfnamefont {G.~J.}\ \bibnamefont
  {Dick}},\ }\href@noop {} {\bibfield  {journal} {\bibinfo  {journal} {Proc.
  19th Annu. Precise Time Time Interval}\ }\textbf {\bibinfo {volume} {19}},\
  \bibinfo {pages} {133} (\bibinfo {year} {1987})}\BibitemShut {NoStop}%
\bibitem [{\citenamefont {Cheinet}\ \emph {et~al.}(2008)\citenamefont
  {Cheinet}, \citenamefont {Canuel}, \citenamefont {Santos}, \citenamefont
  {Gauguet}, \citenamefont {Yver-Leduc},\ and\ \citenamefont
  {Landragin}}]{Cheinet_2008}%
  \BibitemOpen
  \bibfield  {author} {\bibinfo {author} {\bibfnamefont {P.}~\bibnamefont
  {Cheinet}}, \bibinfo {author} {\bibfnamefont {B.}~\bibnamefont {Canuel}},
  \bibinfo {author} {\bibfnamefont {F.~P.~D.}\ \bibnamefont {Santos}}, \bibinfo
  {author} {\bibfnamefont {A.}~\bibnamefont {Gauguet}}, \bibinfo {author}
  {\bibfnamefont {F.}~\bibnamefont {Yver-Leduc}}, \ and\ \bibinfo {author}
  {\bibfnamefont {A.}~\bibnamefont {Landragin}},\ }\href {\doibase
  10.1109/tim.2007.915148} {\bibfield  {journal} {\bibinfo  {journal} {{IEEE}
  Transactions on Instrumentation and Measurement}\ }\textbf {\bibinfo {volume}
  {57}},\ \bibinfo {pages} {1141} (\bibinfo {year} {2008})}\BibitemShut
  {NoStop}%
\bibitem [{\citenamefont {Bize}(2001)}]{BizePhD}%
  \BibitemOpen
  \bibfield  {author} {\bibinfo {author} {\bibfnamefont {S.}~\bibnamefont
  {Bize}},\ }\emph {\bibinfo {title} {Tests fondamentaux {\`a} l'aide
  d'horloges {\`a} atomes froids de rubidium et de c{\'e}sium}},\ \href
  {https://tel.archives-ouvertes.fr/tel-00000981} {\bibinfo {type} {Theses}},\
  \bibinfo  {school} {Universit{\'e} Pierre et Marie Curie - Paris VI}
  (\bibinfo {year} {2001}),\ \bibinfo {note} {[page 49, Eq.(4.21)]}\BibitemShut
  {NoStop}%
\bibitem [{\citenamefont {Meunier}\ \emph {et~al.}(2014)\citenamefont
  {Meunier}, \citenamefont {Dutta}, \citenamefont {Geiger}, \citenamefont
  {Guerlin}, \citenamefont {Alzar},\ and\ \citenamefont
  {Landragin}}]{Meunier_2014}%
  \BibitemOpen
  \bibfield  {author} {\bibinfo {author} {\bibfnamefont {M.}~\bibnamefont
  {Meunier}}, \bibinfo {author} {\bibfnamefont {I.}~\bibnamefont {Dutta}},
  \bibinfo {author} {\bibfnamefont {R.}~\bibnamefont {Geiger}}, \bibinfo
  {author} {\bibfnamefont {C.}~\bibnamefont {Guerlin}}, \bibinfo {author}
  {\bibfnamefont {C.~L.~G.}\ \bibnamefont {Alzar}}, \ and\ \bibinfo {author}
  {\bibfnamefont {A.}~\bibnamefont {Landragin}},\ }\href {\doibase
  10.1103/physreva.90.063633} {\bibfield  {journal} {\bibinfo  {journal}
  {Physical Review A}\ }\textbf {\bibinfo {volume} {90}},\ \bibinfo {pages}
  {063633} (\bibinfo {year} {2014})}\BibitemShut {NoStop}%
\bibitem [{\citenamefont {L{\'{e}}v{\`{e}}que}\ \emph
  {et~al.}(2010)\citenamefont {L{\'{e}}v{\`{e}}que}, \citenamefont {Gauguet},
  \citenamefont {Chaibi},\ and\ \citenamefont {Landragin}}]{L_v_que_2010}%
  \BibitemOpen
  \bibfield  {author} {\bibinfo {author} {\bibfnamefont {T.}~\bibnamefont
  {L{\'{e}}v{\`{e}}que}}, \bibinfo {author} {\bibfnamefont {A.}~\bibnamefont
  {Gauguet}}, \bibinfo {author} {\bibfnamefont {W.}~\bibnamefont {Chaibi}}, \
  and\ \bibinfo {author} {\bibfnamefont {A.}~\bibnamefont {Landragin}},\ }\href
  {\doibase 10.1007/s00340-010-4082-y} {\bibfield  {journal} {\bibinfo
  {journal} {Applied Physics B}\ }\textbf {\bibinfo {volume} {101}},\ \bibinfo
  {pages} {723} (\bibinfo {year} {2010})}\BibitemShut {NoStop}%
\bibitem [{\citenamefont {Merlet}\ \emph {et~al.}(2009)\citenamefont {Merlet},
  \citenamefont {Gouët}, \citenamefont {Bodart}, \citenamefont {Clairon},
  \citenamefont {Landragin}, \citenamefont {Santos},\ and\ \citenamefont
  {Rouchon}}]{Merlet_2009}%
  \BibitemOpen
  \bibfield  {author} {\bibinfo {author} {\bibfnamefont {S.}~\bibnamefont
  {Merlet}}, \bibinfo {author} {\bibfnamefont {J.~L.}\ \bibnamefont {Gouët}},
  \bibinfo {author} {\bibfnamefont {Q.}~\bibnamefont {Bodart}}, \bibinfo
  {author} {\bibfnamefont {A.}~\bibnamefont {Clairon}}, \bibinfo {author}
  {\bibfnamefont {A.}~\bibnamefont {Landragin}}, \bibinfo {author}
  {\bibfnamefont {F.~P.~D.}\ \bibnamefont {Santos}}, \ and\ \bibinfo {author}
  {\bibfnamefont {P.}~\bibnamefont {Rouchon}},\ }\href {\doibase
  10.1088/0026-1394/46/1/011} {\bibfield  {journal} {\bibinfo  {journal}
  {Metrologia}\ }\textbf {\bibinfo {volume} {46}},\ \bibinfo {pages} {87}
  (\bibinfo {year} {2009})}\BibitemShut {NoStop}%
\bibitem [{\citenamefont {Hu}\ \emph {et~al.}(2013)\citenamefont {Hu},
  \citenamefont {Sun}, \citenamefont {Duan}, \citenamefont {Zhou},
  \citenamefont {Chen}, \citenamefont {Zhan}, \citenamefont {Zhang},\ and\
  \citenamefont {Luo}}]{Hu2013}%
  \BibitemOpen
  \bibfield  {author} {\bibinfo {author} {\bibfnamefont {Z.-K.}\ \bibnamefont
  {Hu}}, \bibinfo {author} {\bibfnamefont {B.-L.}\ \bibnamefont {Sun}},
  \bibinfo {author} {\bibfnamefont {X.-C.}\ \bibnamefont {Duan}}, \bibinfo
  {author} {\bibfnamefont {M.-K.}\ \bibnamefont {Zhou}}, \bibinfo {author}
  {\bibfnamefont {L.-L.}\ \bibnamefont {Chen}}, \bibinfo {author}
  {\bibfnamefont {S.}~\bibnamefont {Zhan}}, \bibinfo {author} {\bibfnamefont
  {Q.-Z.}\ \bibnamefont {Zhang}}, \ and\ \bibinfo {author} {\bibfnamefont
  {J.}~\bibnamefont {Luo}},\ }\href {\doibase 10.1103/PhysRevA.88.043610}
  {\bibfield  {journal} {\bibinfo  {journal} {Phys. Rev. A}\ }\textbf {\bibinfo
  {volume} {88}},\ \bibinfo {pages} {043610} (\bibinfo {year}
  {2013})}\BibitemShut {NoStop}%
\bibitem [{\citenamefont {Gouët}\ \emph {et~al.}(2008)\citenamefont {Gouët},
  \citenamefont {Mehlstäubler}, \citenamefont {Kim}, \citenamefont {Merlet},
  \citenamefont {Clairon}, \citenamefont {Landragin},\ and\ \citenamefont
  {Santos}}]{Le_Gou_t_2008}%
  \BibitemOpen
  \bibfield  {author} {\bibinfo {author} {\bibfnamefont {J.~L.}\ \bibnamefont
  {Gouët}}, \bibinfo {author} {\bibfnamefont {T.}~\bibnamefont
  {Mehlstäubler}}, \bibinfo {author} {\bibfnamefont {J.}~\bibnamefont {Kim}},
  \bibinfo {author} {\bibfnamefont {S.}~\bibnamefont {Merlet}}, \bibinfo
  {author} {\bibfnamefont {A.}~\bibnamefont {Clairon}}, \bibinfo {author}
  {\bibfnamefont {A.}~\bibnamefont {Landragin}}, \ and\ \bibinfo {author}
  {\bibfnamefont {F.~P.~D.}\ \bibnamefont {Santos}},\ }\href {\doibase
  10.1007/s00340-008-3088-1} {\bibfield  {journal} {\bibinfo  {journal}
  {Applied Physics B}\ }\textbf {\bibinfo {volume} {92}},\ \bibinfo {pages}
  {133} (\bibinfo {year} {2008})}\BibitemShut {NoStop}%
\bibitem [{\citenamefont {Gouët}\ \emph {et~al.}(2007)\citenamefont {Gouët},
  \citenamefont {Cheinet}, \citenamefont {Kim}, \citenamefont {Holleville},
  \citenamefont {Clairon}, \citenamefont {Landragin},\ and\ \citenamefont
  {Santos}}]{Le_Gou_t_2007}%
  \BibitemOpen
  \bibfield  {author} {\bibinfo {author} {\bibfnamefont {J.~L.}\ \bibnamefont
  {Gouët}}, \bibinfo {author} {\bibfnamefont {P.}~\bibnamefont {Cheinet}},
  \bibinfo {author} {\bibfnamefont {J.}~\bibnamefont {Kim}}, \bibinfo {author}
  {\bibfnamefont {D.}~\bibnamefont {Holleville}}, \bibinfo {author}
  {\bibfnamefont {A.}~\bibnamefont {Clairon}}, \bibinfo {author} {\bibfnamefont
  {A.}~\bibnamefont {Landragin}}, \ and\ \bibinfo {author} {\bibfnamefont
  {F.~P.~D.}\ \bibnamefont {Santos}},\ }\href {\doibase
  10.1140/epjd/e2007-00218-2} {\bibfield  {journal} {\bibinfo  {journal} {The
  European Physical Journal D}\ }\textbf {\bibinfo {volume} {44}},\ \bibinfo
  {pages} {419} (\bibinfo {year} {2007})}\BibitemShut {NoStop}%
\bibitem [{\citenamefont {Canuel}\ \emph {et~al.}(2014)\citenamefont {Canuel},
  \citenamefont {Amand}, \citenamefont {Bertoldi}, \citenamefont {Chaibi},
  \citenamefont {Geiger}, \citenamefont {Gillot}, \citenamefont {Landragin},
  \citenamefont {Merzougui}, \citenamefont {Riou}, \citenamefont {Schmid},\
  and\ \citenamefont {Bouyer}}]{Canuel_2014}%
  \BibitemOpen
  \bibfield  {author} {\bibinfo {author} {\bibfnamefont {B.}~\bibnamefont
  {Canuel}}, \bibinfo {author} {\bibfnamefont {L.}~\bibnamefont {Amand}},
  \bibinfo {author} {\bibfnamefont {A.}~\bibnamefont {Bertoldi}}, \bibinfo
  {author} {\bibfnamefont {W.}~\bibnamefont {Chaibi}}, \bibinfo {author}
  {\bibfnamefont {R.}~\bibnamefont {Geiger}}, \bibinfo {author} {\bibfnamefont
  {J.}~\bibnamefont {Gillot}}, \bibinfo {author} {\bibfnamefont
  {A.}~\bibnamefont {Landragin}}, \bibinfo {author} {\bibfnamefont
  {M.}~\bibnamefont {Merzougui}}, \bibinfo {author} {\bibfnamefont
  {I.}~\bibnamefont {Riou}}, \bibinfo {author} {\bibfnamefont {S.}~\bibnamefont
  {Schmid}}, \ and\ \bibinfo {author} {\bibfnamefont {P.}~\bibnamefont
  {Bouyer}},\ }\href {\doibase 10.1051/e3sconf/20140401004} {\bibfield
  {journal} {\bibinfo  {journal} {E3S Web of Conferences}\ }\textbf {\bibinfo
  {volume} {4}},\ \bibinfo {pages} {01004} (\bibinfo {year}
  {2014})}\BibitemShut {NoStop}%
\bibitem [{\citenamefont {Quessada}\ \emph {et~al.}(2003)\citenamefont
  {Quessada}, \citenamefont {Kovacich}, \citenamefont {Courtillot},
  \citenamefont {Clairon}, \citenamefont {Santarelli},\ and\ \citenamefont
  {Lemonde}}]{quessada2003}%
  \BibitemOpen
  \bibfield  {author} {\bibinfo {author} {\bibfnamefont {A.}~\bibnamefont
  {Quessada}}, \bibinfo {author} {\bibfnamefont {R.~P.}\ \bibnamefont
  {Kovacich}}, \bibinfo {author} {\bibfnamefont {I.}~\bibnamefont
  {Courtillot}}, \bibinfo {author} {\bibfnamefont {A.}~\bibnamefont {Clairon}},
  \bibinfo {author} {\bibfnamefont {G.}~\bibnamefont {Santarelli}}, \ and\
  \bibinfo {author} {\bibfnamefont {P.}~\bibnamefont {Lemonde}},\ }\href
  {\doibase https://doi.org/10.1088/1464-4266/5/2/373} {\bibfield  {journal}
  {\bibinfo  {journal} {J. Opt. B: Quantum Semiclassical Opt.}\ }\textbf
  {\bibinfo {volume} {5}},\ \bibinfo {pages} {S150} (\bibinfo {year}
  {2003})}\BibitemShut {NoStop}%
\bibitem [{\citenamefont {Nicholson}\ \emph {et~al.}(2012)\citenamefont
  {Nicholson}, \citenamefont {Martin}, \citenamefont {Williams}, \citenamefont
  {Bloom}, \citenamefont {Bishof}, \citenamefont {Swallows}, \citenamefont
  {Campbell},\ and\ \citenamefont {Ye}}]{Nicholson2012}%
  \BibitemOpen
  \bibfield  {author} {\bibinfo {author} {\bibfnamefont {T.~L.}\ \bibnamefont
  {Nicholson}}, \bibinfo {author} {\bibfnamefont {M.~J.}\ \bibnamefont
  {Martin}}, \bibinfo {author} {\bibfnamefont {J.~R.}\ \bibnamefont
  {Williams}}, \bibinfo {author} {\bibfnamefont {B.~J.}\ \bibnamefont {Bloom}},
  \bibinfo {author} {\bibfnamefont {M.}~\bibnamefont {Bishof}}, \bibinfo
  {author} {\bibfnamefont {M.~D.}\ \bibnamefont {Swallows}}, \bibinfo {author}
  {\bibfnamefont {S.~L.}\ \bibnamefont {Campbell}}, \ and\ \bibinfo {author}
  {\bibfnamefont {J.}~\bibnamefont {Ye}},\ }\href {\doibase
  https://doi.org/10.1103/PhysRevLett.109.230801} {\bibfield  {journal}
  {\bibinfo  {journal} {Phys. Rev. Lett.}\ }\textbf {\bibinfo {volume} {109}},\
  \bibinfo {pages} {230801} (\bibinfo {year} {2012})}\BibitemShut {NoStop}%
\bibitem [{\citenamefont {Koller}\ \emph {et~al.}(2017)\citenamefont {Koller},
  \citenamefont {Grotti}, \citenamefont {Vogt}, \citenamefont {Al-Masoudi},
  \citenamefont {D\"orscher}, \citenamefont {H\"afner}, \citenamefont {Sterr},\
  and\ \citenamefont {Lisdat}}]{koller2017}%
  \BibitemOpen
  \bibfield  {author} {\bibinfo {author} {\bibfnamefont {S.}~\bibnamefont
  {Koller}}, \bibinfo {author} {\bibfnamefont {J.}~\bibnamefont {Grotti}},
  \bibinfo {author} {\bibfnamefont {S.}~\bibnamefont {Vogt}}, \bibinfo {author}
  {\bibfnamefont {A.}~\bibnamefont {Al-Masoudi}}, \bibinfo {author}
  {\bibfnamefont {S.}~\bibnamefont {D\"orscher}}, \bibinfo {author}
  {\bibfnamefont {S.}~\bibnamefont {H\"afner}}, \bibinfo {author}
  {\bibfnamefont {U.}~\bibnamefont {Sterr}}, \ and\ \bibinfo {author}
  {\bibfnamefont {C.}~\bibnamefont {Lisdat}},\ }\href {\doibase
  https://doi.org/10.1103/PhysRevLett.118.073601} {\bibfield  {journal}
  {\bibinfo  {journal} {Phys. Rev. Lett.}\ }\textbf {\bibinfo {volume} {118}},\
  \bibinfo {pages} {073601} (\bibinfo {year} {2017})}\BibitemShut {NoStop}%
\bibitem [{\citenamefont {Cheinet}(2006)}]{CheinetPhD}%
  \BibitemOpen
  \bibfield  {author} {\bibinfo {author} {\bibfnamefont {P.}~\bibnamefont
  {Cheinet}},\ }\emph {\bibinfo {title} {Conception and realisation of a cold
  atom gravimeter}},\ \href {https://tel.archives-ouvertes.fr/tel-00070861}
  {\bibinfo {type} {Theses}},\ \bibinfo  {school} {Universit{\'e} Pierre et
  Marie Curie - Paris VI} (\bibinfo {year} {2006}),\ \bibinfo {note} {the
  correction in the main text corresponds to Eq.(2.45) on page 38 with $T$
  defined as the time elapsed between the centre of adjacent
  pulses.}\BibitemShut {Stop}%
\bibitem [{\citenamefont {Riou}\ \emph {et~al.}(2017)\citenamefont {Riou},
  \citenamefont {Mielec}, \citenamefont {Lefèvre}, \citenamefont {Prevedelli},
  \citenamefont {Landragin}, \citenamefont {Bouyer}, \citenamefont {Bertoldi},
  \citenamefont {Geiger},\ and\ \citenamefont {Canuel}}]{RiouMielec2017}%
  \BibitemOpen
  \bibfield  {author} {\bibinfo {author} {\bibfnamefont {I.}~\bibnamefont
  {Riou}}, \bibinfo {author} {\bibfnamefont {N.}~\bibnamefont {Mielec}},
  \bibinfo {author} {\bibfnamefont {G.}~\bibnamefont {Lefèvre}}, \bibinfo
  {author} {\bibfnamefont {M.}~\bibnamefont {Prevedelli}}, \bibinfo {author}
  {\bibfnamefont {A.}~\bibnamefont {Landragin}}, \bibinfo {author}
  {\bibfnamefont {P.}~\bibnamefont {Bouyer}}, \bibinfo {author} {\bibfnamefont
  {A.}~\bibnamefont {Bertoldi}}, \bibinfo {author} {\bibfnamefont
  {R.}~\bibnamefont {Geiger}}, \ and\ \bibinfo {author} {\bibfnamefont
  {B.}~\bibnamefont {Canuel}},\ }\href
  {http://stacks.iop.org/0953-4075/50/i=15/a=155002} {\bibfield  {journal}
  {\bibinfo  {journal} {Journal of Physics B: Atomic, Molecular and Optical
  Physics}\ }\textbf {\bibinfo {volume} {50}},\ \bibinfo {pages} {155002}
  (\bibinfo {year} {2017})}\BibitemShut {NoStop}%
\bibitem [{\citenamefont {Hamilton}\ \emph {et~al.}(2015)\citenamefont
  {Hamilton}, \citenamefont {Jaffe}, \citenamefont {Brown}, \citenamefont
  {Maisenbacher}, \citenamefont {Estey},\ and\ \citenamefont
  {M\"uller}}]{Hamilton2015}%
  \BibitemOpen
  \bibfield  {author} {\bibinfo {author} {\bibfnamefont {P.}~\bibnamefont
  {Hamilton}}, \bibinfo {author} {\bibfnamefont {M.}~\bibnamefont {Jaffe}},
  \bibinfo {author} {\bibfnamefont {J.~M.}\ \bibnamefont {Brown}}, \bibinfo
  {author} {\bibfnamefont {L.}~\bibnamefont {Maisenbacher}}, \bibinfo {author}
  {\bibfnamefont {B.}~\bibnamefont {Estey}}, \ and\ \bibinfo {author}
  {\bibfnamefont {H.}~\bibnamefont {M\"uller}},\ }\href {\doibase
  10.1103/PhysRevLett.114.100405} {\bibfield  {journal} {\bibinfo  {journal}
  {Phys. Rev. Lett.}\ }\textbf {\bibinfo {volume} {114}},\ \bibinfo {pages}
  {100405} (\bibinfo {year} {2015})}\BibitemShut {NoStop}%
\bibitem [{\citenamefont {Decamps}(2016)}]{Decamps2016}%
  \BibitemOpen
  \bibfield  {author} {\bibinfo {author} {\bibfnamefont {B.}~\bibnamefont
  {Decamps}},\ }\emph {\bibinfo {title} {Atom interferometry: experiments with
  electromagnetic interactions and design of a Bose Einstein condensate
  setup}},\ \href@noop {} {Ph.D. thesis},\ \bibinfo  {school} {Universit\'e
  Toulouse III - Paul Sabatier} (\bibinfo {year} {2016})\BibitemShut {NoStop}%
\bibitem [{\citenamefont {Clad{\'{e}}}\ \emph {et~al.}(2010)\citenamefont
  {Clad{\'{e}}}, \citenamefont {Plisson}, \citenamefont
  {Guellati-Kh{\'{e}}lifa}, \citenamefont {Nez},\ and\ \citenamefont
  {Biraben}}]{Clad_2010}%
  \BibitemOpen
  \bibfield  {author} {\bibinfo {author} {\bibfnamefont {P.}~\bibnamefont
  {Clad{\'{e}}}}, \bibinfo {author} {\bibfnamefont {T.}~\bibnamefont
  {Plisson}}, \bibinfo {author} {\bibfnamefont {S.}~\bibnamefont
  {Guellati-Kh{\'{e}}lifa}}, \bibinfo {author} {\bibfnamefont {F.}~\bibnamefont
  {Nez}}, \ and\ \bibinfo {author} {\bibfnamefont {F.}~\bibnamefont
  {Biraben}},\ }\href {\doibase 10.1140/epjd/e2010-00198-0} {\bibfield
  {journal} {\bibinfo  {journal} {The European Physical Journal D}\ }\textbf
  {\bibinfo {volume} {59}},\ \bibinfo {pages} {349} (\bibinfo {year}
  {2010})}\BibitemShut {NoStop}%
\end{thebibliography}%
\bibliographystyle{apsrev4-1}
\label{sec:bibliography}


\appendix
\section{Definition of the pulse shapes}
\label{sec:pulsedef}

We define the time-dependent Rabi frequency as
\begin{equation}
\Omega(t)=\Omega_0 f(t),
\end{equation}
with $\Omega_0$ the peak Rabi frequency. The pulses are defined by the function $f(t)$ with maximal amplitude $1$.

The GSinc pulse is defined as 
\begin{equation}
f(t) = \text{sinc}(\pi t / t_1) \times e^{- \frac{1}{2} \frac{t^2}{\alpha^2 t_1^2}}
\end{equation}
With $\text{sinc}(x)=\sin(x) / x$,  $t_1$ the time of the first zero of the sinc, $\alpha t_1$ the standard deviation  of the gaussian modulation. The total  pulse duration is defined as $2 n t_1$. 
In Fig.~\ref{fig:Homega} the parameters of the GSinc pulse are $n=6$ and $\alpha=2.4$. \\
The GFlat pulse is even and defined as 
\begin{equation}
f(t) = \left\{
		\begin{array}{ll}
			1 \quad \text{if} \quad t < t_0\\
			e^{-\frac{1}{2} \frac{(t-t_0)^2}{r^2 t_0^2}} \quad \text{if} \quad t > t_0\\
		\end{array}
	   \right.
\end{equation}
where $t_0$ is the half length of the plateau, and $r t_0$ is the standard deviation of the gaussian. The total pulse duration is defined as $2t_0 + 2 n r t_0$.
In the main text, we consider GFlat pulses with $r=1$ and $n=6$.

\begin{figure}[!h]
 \includegraphics[width= 0.9\linewidth]{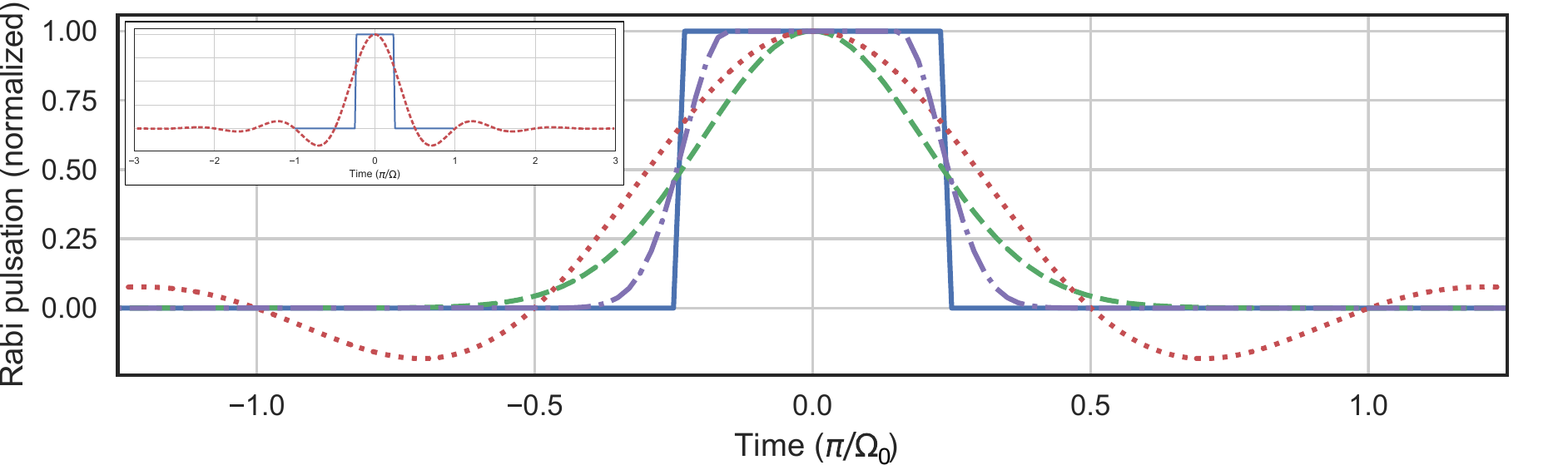}
\caption{\label{fig:detuning_selectivity}
Illustration of the different pulse shapes considered in this article: rectangular (plain blue line), gaussian (green dashed), GFlat (violet dot-dashed), GSinc (dotted red). Note that the peak Rabi frequency is kept constant for all pulse shapes.
 For ease of illustration, we have cropped the GSinc pulse to its center part in the main panel. The inset shows the full GSinc pulse shape.
 The time axis is in units of the inverse Rabi frequency.
}
\end{figure}

The pulse shapes are illustrated in Fig.~\ref{fig:detuning_selectivity}.

\begin{figure}[!h]
 \includegraphics[width= 0.8\linewidth]{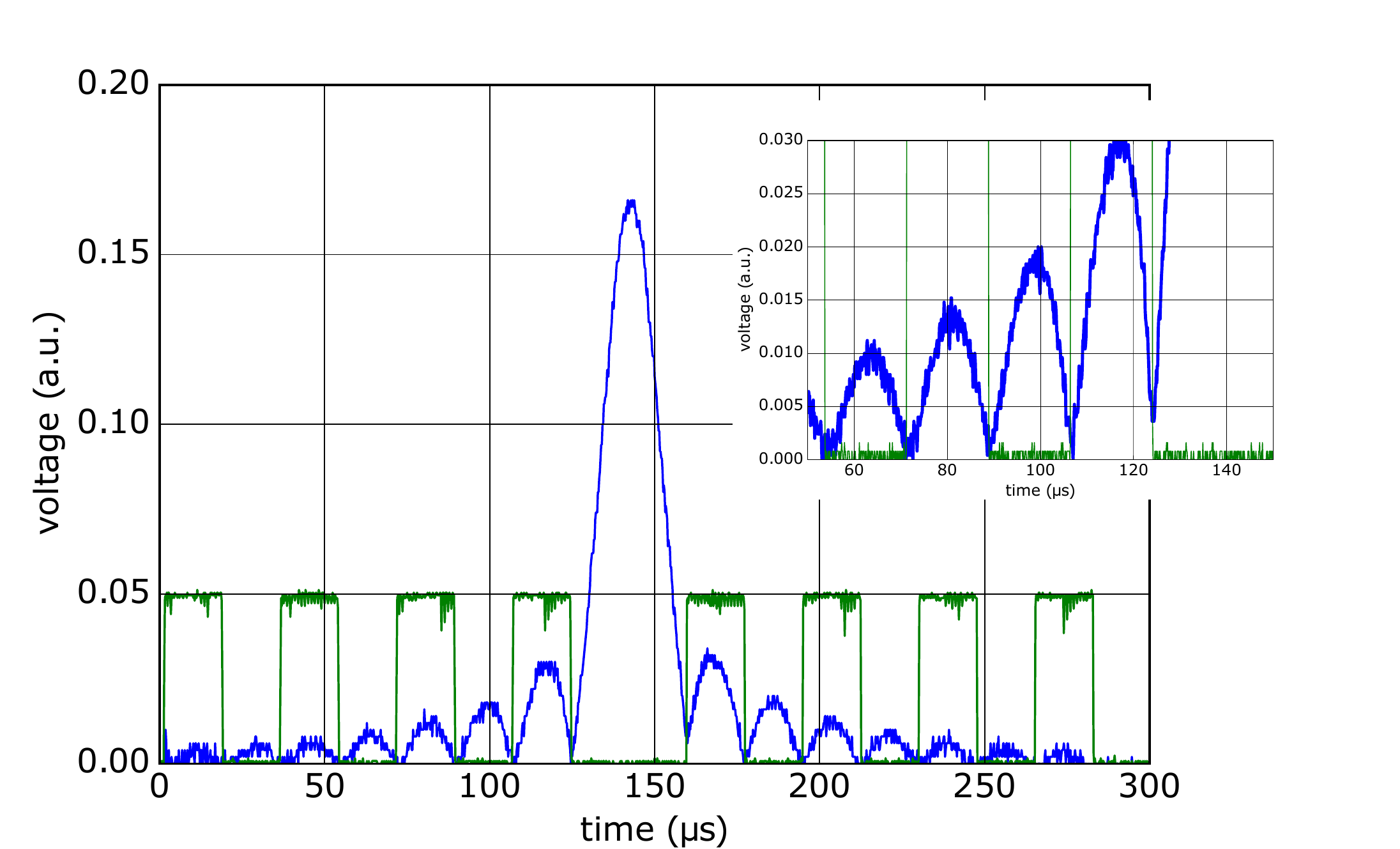}
\caption{\label{fig:time_trace_sinc} Time trace of the sinc pulse with 8 zeros on each side of the maximum. The blue line shows the voltage recorded by the photodiode which monitors the power of the Raman beam. The green trace shows the digital signal which triggers phase jumps of $\pi$ applied to the phase lock loop. The inset is a zoom on the zeros of the power and on the phase jumps.
}
\end{figure}

In section \ref{sec:selectivity} we study experimentally several Sinc pulse shapes with different number of zeros on each side of the maximum.
As an illustration of  implementation of such pulses, a time trace of a Sinc pulse with 8 zeros on each side of the maximum is shown in Fig.\ref{fig:time_trace_sinc}.

\section{Details on the qualitative study of the  influence of the pulse shape on the transfer function}
\label{sec:qualitative_interpretation}

\begin{figure}[!bth]
 \includegraphics[width= \linewidth]{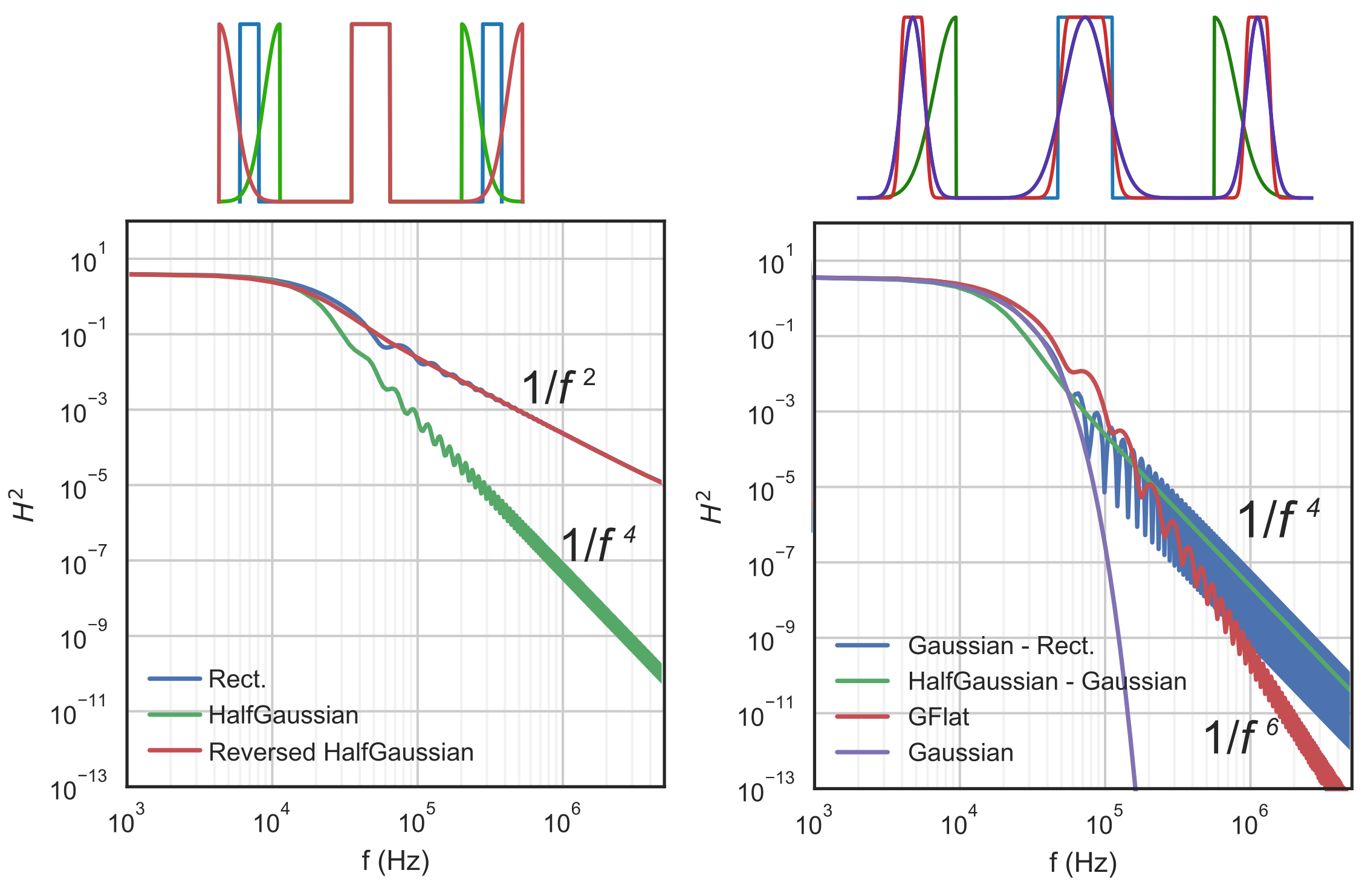}
\caption{\label{fig:tf_decay}
Illustration of the  behavior of $|H(2\pi f)|^2$ at  high frequency using different sequences of pulses.
The calculations are for a 3 pulse interferometer with $12$~kHz peak Rabi frequency and $T=20$~ms, using different pulse shapes. The top panel shows the considered pulse sequences. 
Left: Evolution from the $1/f^2$ scaling to the $1/f^4$ scaling, which occurs when smoothing the outer parts of the interferometer pulses, i.e.  the beginning of the first $\pi/2$ pulse and  the end  of the last $\pi/2$ pulse.
Right: Evolution from the $1/f^4$ scaling to even faster decays when smoothing the inner parts of the interferometer pulses.
}
\end{figure}

The high frequency behavior of the transfer function can be qualitatively understood from the pulse shape according to the position of the pulses in the interferometric sequence.
We recall that the transfer function is $|H(\omega)|^2=\omega^2 |G(\omega)|^2 $, where $G(\omega)$ is the Fourier transform of $g(t)$, which is itself the sine of the integral of the time-dependent Rabi frequency, see Eq.\eqref{eqn:integral_g}.
We  define $I(t) = \int_{-\infty}^t\Omega(u)du$.

Our first observation, illustrated in Fig.~\ref{fig:tf_decay}(left), is that a decay faster than $1/f^2$ can be obtained by smoothing the beginning of the first $\pi/2$ and the end of the last $\pi/2$ pulses. 
At these points in time, where $I(t)\simeq 0$,  the sensitivity function can be Taylor-expanded as $g(t) \simeq I(t) + O(I(t)^3)$.
A rectangular pulse results in a triangular form of $I(t)$, giving rise to a $1/f^2$ dependence in $|G(2\pi f)|$ and hence to a $1/f^2$ dependence in $|H(2\pi f)|^2$. In contrast,  smooth pulses are characterized by a slower growth of $I(t)$ and hence a faster decay of $|H(2\pi f)|^2$.
This is illustrated in Fig.~\ref{fig:tf_decay}(left) by calculating the transfer function using half Gaussian pulses for the $\pi/2$ pulses and a rectangular $\pi$ pulse.

Evolution from $1/f^4$ to a faster decay is governed by the end of the first $\pi/2$ pulse, the beginning of the last $\pi/2$ pulse, and the beginning and end of the central $\pi$ pulse. 
At these positions, $I(t)\simeq\pi/2$, and the sensitivity function can be approximated by $g(t) \simeq 1 - \frac{1}{2} I(t)^2 + O(I(t)^4)$. 
Here the leading order of the time-dependence is quadratic, which explains why the influence of this part of the pulses has a weaker influence on the high frequency behavior.
The rectangular $\pi$-pulse, for example, results in a parabolic shape of $I(t)$, yielding a $1/f^4$ dependence of $|H|^2$.
Fig.~\ref{fig:tf_decay} (right) illustrates  the need to smooth these parts of the pulses in order to obtain a decay faster than $1/f^4$ in the transfer function.

\section{Transfer function for an atom interferometer in an optical cavity}
\label{sec:cavityTF}

\begin{figure}[!bth]
 \includegraphics[width= 0.8\linewidth]{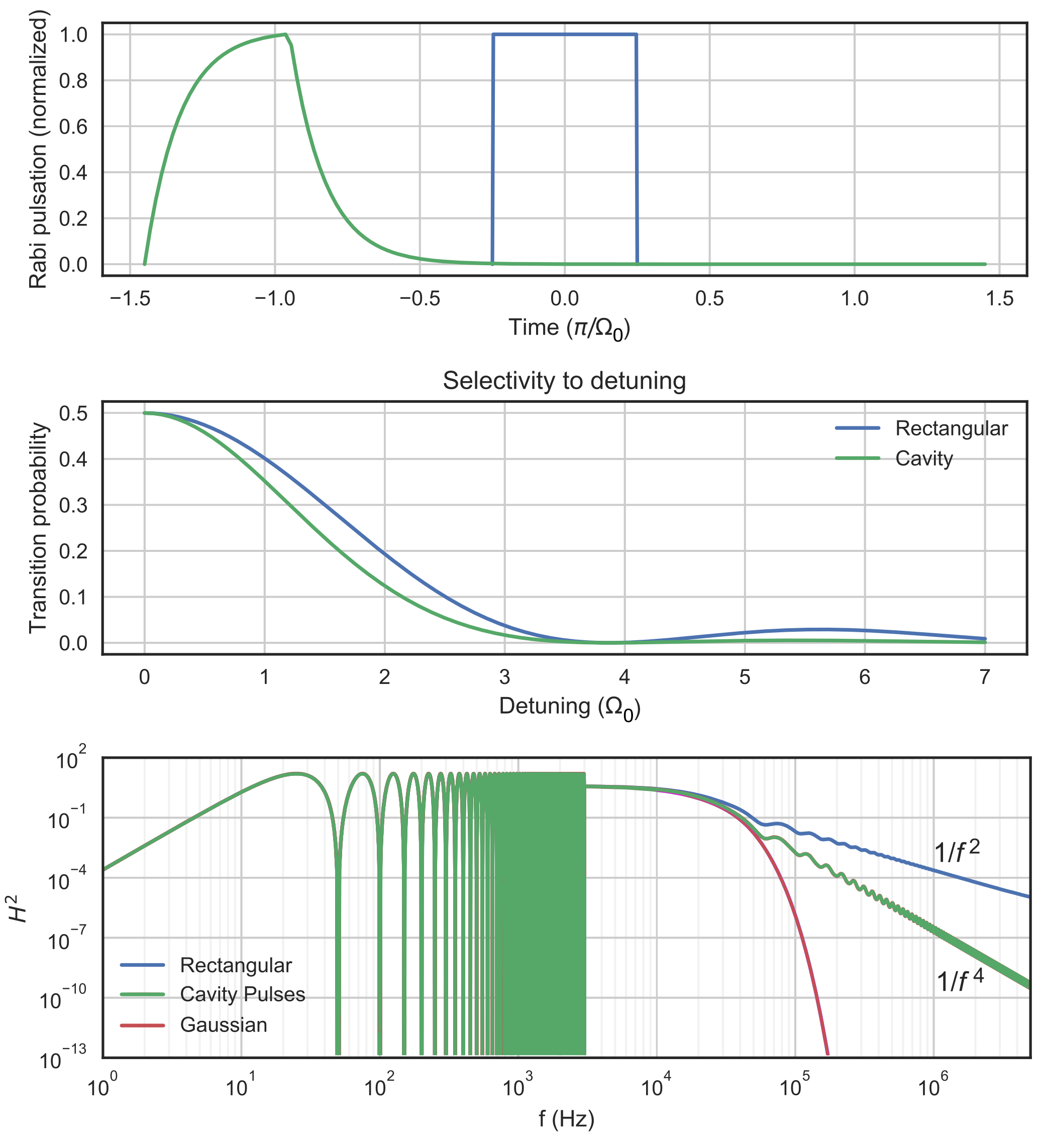}
\caption{\label{fig:cavityTF}
(Top) Shape of a cavity-like pulse and (Middle) selectivity to detuning for a $\pi/2$ pulse. Here $\tau_{cav}=5 \ \mu$s.  We show the shape of a rectangular pulse for comparison.
Bottom: $|H(2\pi f)|^2$ in a 3 pulse interferometer with $12$~kHz peak Rabi frequency and $T=20$~ms. We show here again the response of the rectangular and Gaussian pulses for the ease of comparison.
}
\end{figure}

We present in Fig.~\ref{fig:cavityTF} the temporal shape (top), the velocity selectivity (middle) and the transfer function for a pulse shape resembling the response of an optical cavity. We assumed an intensity build up time of $\tau_{cav}=5 \ \mu$s.  Compared with rectangular pulses (blue), cavity pulses is more selective to detuning but rejects better the high-frequency laser phase noise.  

\end{document}